\documentclass[aps,pre,superscriptaddress,twocolumn,amsmath,amssymb,showpacs]{revtex4-1}
\usepackage{amsmath}
\usepackage{amssymb}
\usepackage{graphicx}
\usepackage{float}
\usepackage{subfigure}
\usepackage{color}
\usepackage{hyperref}
\usepackage{comment}
\usepackage[utf8x]{inputenc}

\begin{document}
\title{One-point height fluctuations and two-point correlators of $(2+1)$ cylindrical KPZ systems}
\author{Ismael S. S. Carrasco}
\email{ismael.carrasco@unb.br}
\affiliation{University of Brasilia, International Center of Physics, Institute of Physics, 70910-900, Brasilia, Federal District, Brazil}
\author{Tiago J. Oliveira}
\email{tiago@ufv.br}
\affiliation{Departamento de F\'isica, Universidade Federal de Vi\c cosa, 36570-900, Vi\c cosa, MG, Brazil}
\date{\today}

\begin{abstract}
While the 1-point height distributions (HDs) and 2-point covariances of $(2+1)$ KPZ systems have been investigated in several recent works for flat and spherical geometries, for the cylindrical one the HD was analyzed for few models and nothing is known about the spatial and temporal covariances. Here, we report results for these quantities, obtained from extensive numerical simulations of discrete KPZ models, for three different setups yielding cylindrical growth. Beyond demonstrating the universality of the HD and covariances, our results reveal other interesting features of this geometry. For example, the spatial covariances measured along the longitudinal and azimuthal directions are different, with the former being quite similar to the curve for flat $(2+1)$ KPZ systems, while the latter resembles the Airy$_2$ covariance of circular $(1+1)$ KPZ interfaces. We also argue (and present numerical evidence) that, in general, the rescaled temporal covariance $\mathcal{A}(t/t_0)$ decays asymptotically as $\mathcal{A}(x) \sim x^{-\bar{\lambda}}$ with an exponent $\bar{\lambda} = \beta + d^*/z$, where $d^*$ is the number of interface sides kept fixed during the growth (being $d^* = 1$ for the systems analyzed here). Overall, these results complete the picture of the main statistics for the $(2+1)$ KPZ class.
\end{abstract}

\maketitle

\section{Introduction}
\label{secIntro}

Since the seminal work by Pr\"ahofer and Spohn \cite{Prahofer2000,*Prahofer2000a}, demonstrating that the asymptotic 1-point (1-pt) height distributions (HDs) for the growth regime (GR) of the one-dimensional [1D or $(1+1)$] polynuclear growth model are given by different probability density functions (pdf's) depending on whether the initial condition (IC) of the growth is a long flat line, a single seed (yielding a droplet-like interface) or stationary, we have witnessed a significant advance in the understanding of 1D Kardar-Parisi-Zhang (KPZ) \cite{KPZ} systems. In fact, motivated by Ref. \cite{Prahofer2000,*Prahofer2000a}, a large number of theoretical \cite{Sasamoto2010,*Amir,*Calabrese2011,*Imamura}, experimental \cite{Takeuchi2010,*Takeuchi2011,*TakeuchiCross} and numerical works \cite{Alves11,*tiago12a,*HealyCross,*silvia17,*Santalla_2015,*Alves18,*Roy,Alves13,Ismael14,HHTake2015} have confirmed that the HDs for the 1D KPZ class are universal, but dependent on the ICs or geometry. More specifically, the asymptotic temporal evolution of the 1-pt height, during the GR, is given by \cite{Krug1992,Prahofer2000,*Prahofer2000a}
\begin{equation}
 h(\vec{x},t) \simeq v_{\infty} t + s_\lambda (\Gamma t)^{\beta} \chi,
\label{eqAnsatz}
\end{equation}
where the asymptotic growth velocity $v_{\infty}$, the signal of the nonlinear coefficient ($\lambda$) in the KPZ equation $s_{\lambda}$, and the amplitude $\Gamma$ are system-dependent parameters; whereas the growth exponent $\beta$ and the pdf's $P(\chi)$ for the random variable $\chi$ are universal. When the growth starts from a flat substrate with \textit{fixed size} $L$ [a single seed, such that $L(t) \sim t$] the HD $P(\chi)$ is given by the Tracy-Widom (TW) distribution from a Gaussian orthogonal [unitary] ensemble (GOE) [(GUE)], while the Baik-Rains \cite{BR} distribution is found for stationary 1D KPZ systems \cite{Prahofer2000,*Prahofer2000a}. As demonstrated in Ref. \cite{Ismael19}, the TW GUE HD is found even when $L(t)$ varies non-linearly in time, provided that it increases faster than the lateral correlation length $\xi \sim t^{1/z}$, where $z$ is the dynamic exponent.

Besides the HDs, the dependence on geometry also manifests in 2-pt correlators. For example, in the GR, the spatial covariance,
\begin{equation}
C_S(r,t) =\langle \tilde{h}(\vec{x},t) \tilde{h}(\vec{x}+\vec{r},t) \rangle \simeq (\Gamma t)^{2\beta} \Psi[A_h r/(\Gamma t)^{2\beta}]
\label{eqCovS}
\end{equation}
has a scaling function $\Psi[s]$ given by different Airy processes for 1D KPZ interfaces with flat and single seed ICs \cite{Prahofer2002,Sasa2005}. In Eq. \ref{eqCovS}, $\tilde{h} = h - \bar{h}$, with $\bar{h}$ being the mean height of the interface at a given time, and $A_h$ is related to the amplitude $A$ of the height-difference correlation function $G_2(r,t) = \langle [h(\vec{x},t) -  h(\vec{x}+\vec{r},t)]^2 \rangle \simeq A r^{2\alpha}$, where $\alpha$ is the roughness exponent. Moreover, the temporal covariance:
\begin{equation}
C_T(t,t_0) = \langle \tilde{h}(\vec{x},t_0) \tilde{h}(\vec{x},t) \rangle \simeq (\Gamma^2 t_0 t)^{\beta}\mathcal{A}(t/t_0)
\label{eqCovT}
\end{equation}
has $\mathcal{A}(y) \sim y^{-\bar{\lambda}}$ for large $y$, with $\bar{\lambda}$ being conjectured to be $\bar{\lambda} = \beta - d/z$ for flat interfaces of dimension $d$ \cite{Kallabis99} and $\bar{\lambda} = \beta$ for (hyper)spherical interfaces (of any $d$) \cite{Singha2005}.

\begin{figure}[!t]
	\centering
	\includegraphics[width=8.5cm]{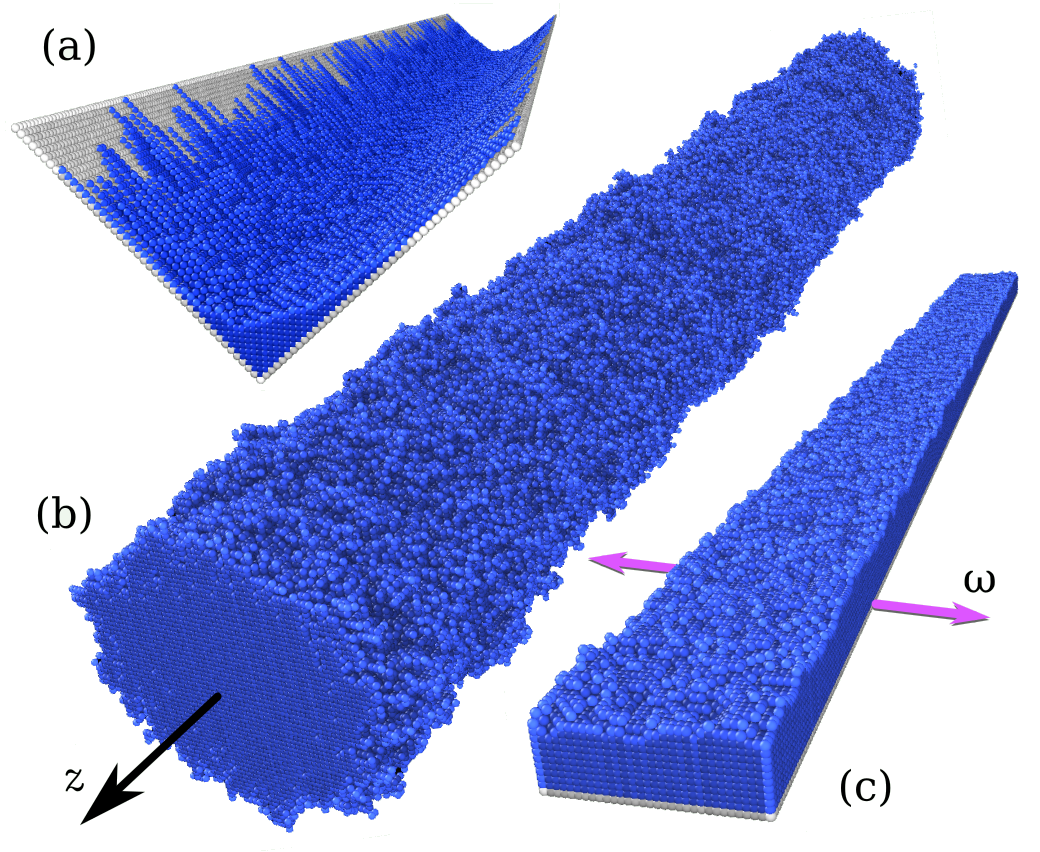}
	\caption{Examples of KPZ interfaces obtained in the three growth setups yielding $(2+1)$-cylindrical geometry: (a) V-shaped groove; (b) Eden cluster grown from a seed-line (on the $z$ axis); and (c) rectangular (flat) substrates where the smaller side enlarges at rate $\omega$. Note that, in all cases, the interface expands in a single direction, while its other side is kept fixed and has a very large size $L_z \gg 1$. The interfaces in (a) and (c) are for the RSOS model.}
	\label{fig1}
\end{figure}

For 2D interfaces, beyond the stationary, the flat ones, and those that expand radially starting from a single seed (spherical geometry), there also exists the interesting situation where the radial growth starts from a long straight line, yielding a cylindrical deposit [see Fig. \ref{fig1}(b)]. The HDs for these ICs were investigated by Halpin-Healy \cite{healy12,healy13}, numerically demonstrating their geometry dependence. Subsequent works have confirmed the universality of the HD for the 2D flat KPZ subclass both numerically \cite{tiago13,Ismael14,Alves14BD} and experimentally \cite{Almeida14,healy14exp,Almeida15,Almeida17}. Moreover, the universality of the HD for the spherical geometry has been verified for models with height restrictions deposited on a corner \cite{tiago13}, and in the growth of $(2+1)$ Eden clusters \cite{tiago13,healy13}, as well as for deposition on enlarging (flat) substrates, whose average lateral sizes increase isotropically as $\langle L_y \rangle = \langle L_z \rangle \sim t$ \cite{Ismael14,HHTake2015}. The same HD is also found when $\langle L_y \rangle = \langle L_z \rangle \sim t^{\gamma}$, provided that $\gamma > 1/z$ \cite{Ismael22}. Furthermore, the spatial covariance for 2D flat KPZ systems has been numerically \cite{healy14exp,Ismael14} and experimentally \cite{healy14exp,Almeida17} studied, while the one for the spherical case was also numerically analyzed in Ref. \cite{Ismael14}. These results strongly suggest that, similarly to the 1D case, the scaling functions $\Psi[s]$ are universal, but dependent on the geometry of the system. The very same conclusion was obtained for the temporal covariances of 2D KPZ models with flat and spherical geometries \cite{Ismael14}. However, to the best of our knowledge, these spatial and temporal correlators were never investigated in the literature for cylindrical KPZ growth.

To address this important issue, we present here a thorough analysis of the HD and covariances for the cylindrical case. Results from extensive numerical simulations of three types of systems are reported: (\textit{i}) models with height restrictions deposited inside long V-shaped grooves (VG); (\textit{ii}) $(2+1)$ Eden clusters growing radially on the cubic lattice from a long straight line; and (\textit{iii}) deposition on rectangular enlarging substrates (ESs) where $L_z \gg 1$ is kept fixed, while $L_y$ expands as $\langle L_y \rangle = L_0 + \omega t$. An illustration of these three growth setups is depicted in Fig. \ref{fig1}. Accounting for the interface anisotropy present in the VG and Eden systems, in all cases, we find cumulants (and ratios of them) for the HDs in striking agreement with the values previously estimated in Refs. \cite{healy12,healy13}, confirming the universality of this HD. Importantly, we demonstrate that the spatial covariances measured in the $y$ and $z$ directions are not identical in these systems, though $\Psi_y[x]$ and $\Psi_z[x]$ are also universal and present intriguing similarities with covariances previously found for other KPZ subclasses. Strong evidence of universality is also found for the temporal covariance, where a general expression (valid for all geometries in any dimension $d$) is conjectured for the exponent $\bar{\lambda}$ related to its asymptotic decay, which is numerically confirmed here for the cylindrical case.

The rest of the paper is organized as follows. In Sec. \ref{secMod}, we define the investigated models and growth setups analyzed, as well as the kinetic Monte Carlo methods used for simulating them. Section \ref{secResWHDs} brings results for the roughness scaling and HDs, while those for the covariances are presented in Sec. \ref{secResCov}. Our final discussions and conclusions are summarized in Sec. \ref{secConc}.

\section{Models}
\label{secMod}

We study the restricted solid-on-solid (RSOS) \cite{kk}, the single step (SS) \cite{ss1} and the Eden model \cite{Eden}, which are paradigmatic systems belonging to the KPZ class \cite{Barabasi}. In the first two models, particles are randomly and sequentially released towards a substrate (which has lateral sizes $L_y \times L_z$ here) and may be reflected back or aggregate at a given site $(i,j)$ depending on the local height difference $\delta h_{ij}= h_{ij}-h_{\partial_{ij}}$, where $\partial_{ij}$ denotes a nearest neighbor (NN) site of $(i,j)$. More specifically, in the RSOS model, the height at a randomly sorted site $(i,j)$ is changed by $h_{ij} \rightarrow h_{ij} + 1$ if the condition $|\delta h_{ij}| \le 1$ is satisfied for all NN's $\partial_{ij}$; while in the SS case, the height is incremented by $h_{ij} \rightarrow h_{ij} + 2$, provided that $|\delta h_{ij}| = 1$, always.

We investigate the RSOS and SS models on two types of substrates: V-shaped grooves and flat rectangular domains whose smaller side ($L_y$) expands. In the former case, deposition is performed on long 1D grooves, of length $L_z$, with periodic boundary conditions (PBCs) in the $z$ direction. These grooves have wedgelike cross sections, such that $h_{ij}(t=0)=|j|$ for all $i=1,\ldots,L_z$ in the RSOS case, where $j=-(L_y-1)/2,\ldots,(L_y-1)/2$ for odd $L_y$. To allow for a SS growth on it, the substrate sites have to satisfy the SS condition $|\delta h_{ij}| = 1$ and, thus, we consider $h_{ij}(t=0)=|j|+[1+(-1)^i]/2$, with even $L_z$. Note that, due to the restrictions in $\delta h_{ij}$, depositions are initially accepted only at the bottom line of the grooves and, as time evolves, the width $\ell_{y,i}(t)$ of the active zone (where aggregation occurs) increases as $\ell_{y,i} \sim t$. We work here with an effectively infinite $L_y$, in the sense that $\ell_{y,i}<L_y$ even for the longest deposition times considered. As usual, the time unity corresponds to the deposition (attempt) of one monolayer (i.e., $L_y L_z$ particles) over the whole substrate, which may be done by depositing only in the area $A(t) = \sum_{i=1}^{L_z} \ell_{y,i}(t)$ of the active zone and updating the time by $t \rightarrow t + 1/A(t)$ after each deposition attempt.

Although the interfaces obtained in the grooves are translation invariant in the $z$ direction, this is not the case in the $y$ direction. So, to obtain interfaces that expand in one direction and whose sites are all statistically equivalent, we investigate the RSOS and SS models also on flat substrates where $L_z$ is large and fixed, while $L_y$ expands linearly in time as $\langle L_y \rangle = L_0 + \omega t$, with PBC in both directions. Flat [$h_{ij}(t=0)=0$] and checkerboard [$h_{ij}(t=0)$ alternating between 0 and 1] ICs are used for the RSOS and SS models, respectively. Following Refs. \cite{Ismael14,Ismael19,Ismael22}, the substrate expansion is performed by stochastically mixing particle deposition, which occurs with probability $P_{dep} = L_y L_z/(L_y L_z+\omega)$, with duplications of randomly chosen columns (in the $z$ direction), occurring with complementary probability: $P_{dup} = \omega/(L_y L_z+\omega)$. After each such event, the time is updated by $t \rightarrow t+1/(L_y L_z+\omega)$.

We also investigate version A of the Eden model, starting from a straight line with $L_z$ occupied sites. This seed line is located on the $z$ axis, which is placed in the middle of a simple cubic lattice with lateral sizes $L \times L \times L_z$. PBC is considered in the $z$ direction, while $L$ is effectively infinite; namely, the diameters of the radially growing clusters are always smaller than $L$ in our analyses. The simulation proceeds as follows: at each time step, one of the $N_p$ sites at the periphery of the cluster [note that $N_p(t=0)=4 L_z$] is randomly chosen and occupied with a new particle, with the time being increased as $t \rightarrow t+1/N_p(t)$.

The simulations on enlarging flat substrates were performed for $\omega=1$ and $L_0=4$. Since we are interested in analyzing the asymptotic fluctuations, for $L_z \rightarrow \infty$, large values of $L_z$ were considered in the simulations for V-shaped grooves and expanding substrates, up to $L_z = 2^{15}$. In the Eden case, however, one has to deal with 3D clusters in the cubic lattice, which would require a prohibitively large amount of RAM memory to be simulated for the same sizes and times considered in the other models. Therefore, the Eden simulations were limited to $L_z = 2^{11}$ and short times. In all cases, the number of samples grown was such that the total number of surface sites considered in the statistics was $\gtrsim 10^{7}$.

\section{Results for the height distributions}
\label{secResWHDs}

In this section, we study the HDs of ($2+1$) cylindrical systems, aiming to demonstrate that they all belongs to the same KPZ subclass and, conversely, to confirm the universality of this HD. We start by recalling that, as pointed out above, the interfaces obtained in the VG case are not translation invariant along the $y$ direction and, as a consequence of this, only the height fluctuations at the central (bottom) line will be analyzed for these systems. Moreover, we remark that Eden clusters growing radially in hypercubic lattices are long-known to acquire anisotropic shapes, because the growth velocity $v(t) = \partial_t \langle h \rangle$ is slightly larger along the lattice directions (see, e.g., Refs. \cite{Alves13} and \cite{tiago13} for discussions on this, respectively, for 2D and 3D clusters starting from a single seed). The same issue appears in the cylindrical clusters analyzed here, since their average height (or radius) grows slightly faster in the $xz$ and $yz$ planes (i.e., in the $\langle 100 \rangle$ directions) than in other radial directions. This is demonstrated in Fig. \ref{FigApendA}(a) of Appendix \ref{secApendA}, where one sees that $v(t)$ is always larger in the $\langle 100 \rangle$ case than in the diagonal planes (i.e., the $\langle 110 \rangle$ directions). Thereby, the clusters' cross sections parallel to the $xy$ plane acquire asymptotic diamond-like shapes, similarly to what happens with Eden clusters on the square lattice, and thus we will analyze their height fluctuations considering only the four statistically equivalent lines in the $xz$ and $yz$ planes (referred to as Eden $\langle 100 \rangle$) and also the four ones in the diagonal planes (Eden $\langle 110 \rangle$).

\begin{figure}[!t]
	\centering
    \includegraphics[width=8.cm]{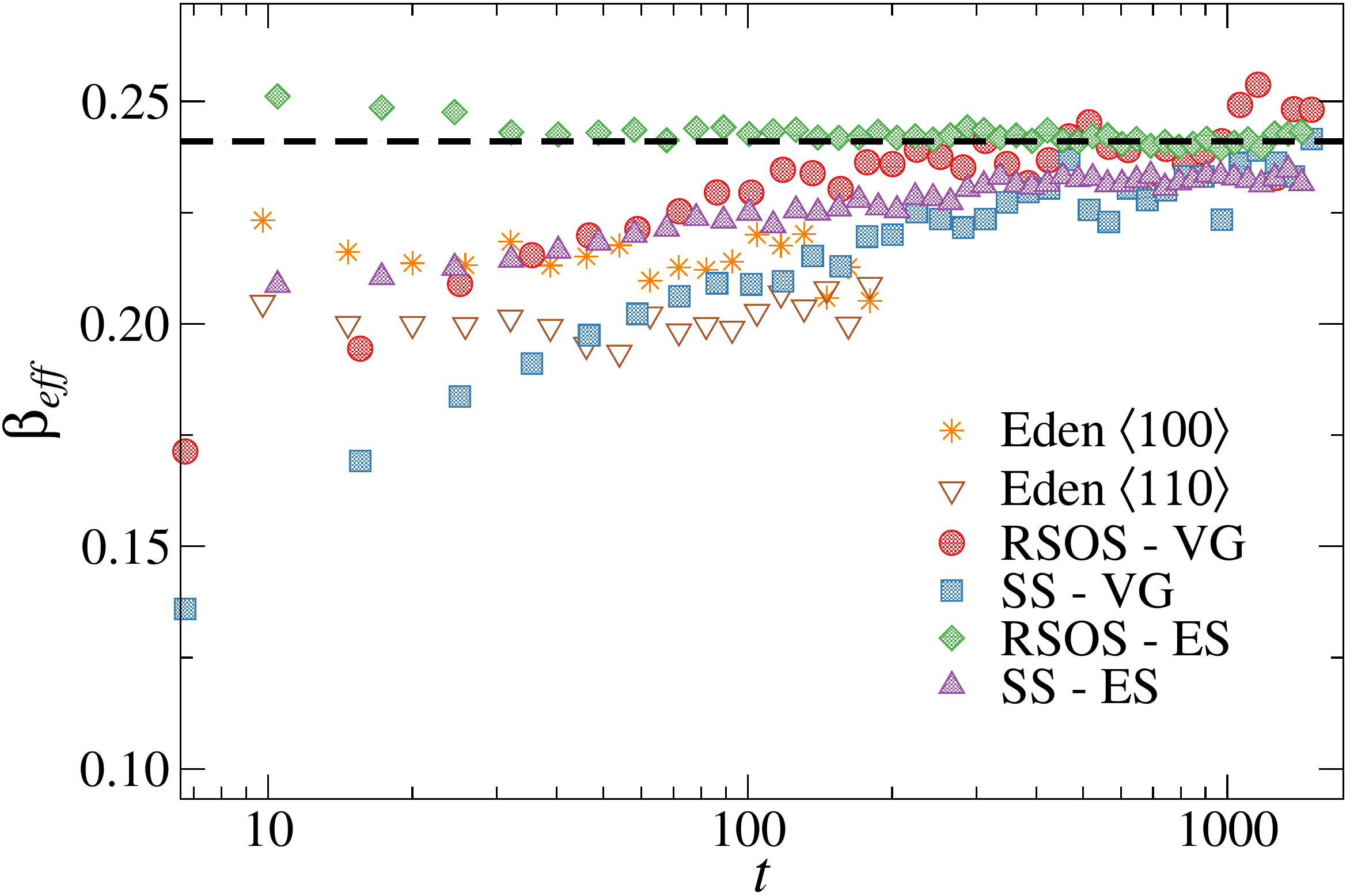}
	\caption{Effective growth exponents, $\beta_{eff}$, versus time, $t$, for the different models and setups analyzed, as shown in the legend. The dashed line indicates the KPZ value $\beta=0.241$.}
	\label{Fig2}
\end{figure}

Before investigating the HDs, it is important to confirm that all systems analyzed here scale with the 2D KPZ exponent $\beta$. The best estimates in the literature for this exponent ($\beta=0.2415(15)$ \cite{Kelling2011}, $\beta=0.2414(2)$ \cite{Kelling2018} and $\beta=0.2399(8)$ \cite{Ismael22}) give an average value $\beta \approx 0.241$, which also agrees with the recent rational conjecture $\beta=7/29$ \cite{tiago22}. Figure \ref{Fig2} shows the temporal evolution of effective growth exponents, $\beta_{eff}$, calculated as the successive slopes in curves of $\ln w^2$ versus $\ln t$, where $w^2 = \langle h^2 \rangle_c = \langle h^2 \rangle  - \langle h \rangle^2$ is the second cumulant of the HDs (whose $n$th cumulant will be denoted here as $\langle h^n \rangle_c$). We may see a clear convergence, and a good agreement at long times in some cases, of the exponents for the RSOS and SS models, for both the VG and ES growth setups, with the expected value. On the other hand, for the Eden model, the effective exponents are still appreciably smaller than $\beta=0.241$ even at the longest times analyzed. This is certainly a consequence of the smaller times simulated in this case, since the Eden exponents are very close to those for the other models at short times, strongly indicating that they shall also converge to the KPZ value when $t \rightarrow \infty$.

\begin{figure}[!t]
	\centering
	\includegraphics[height=3.6cm]{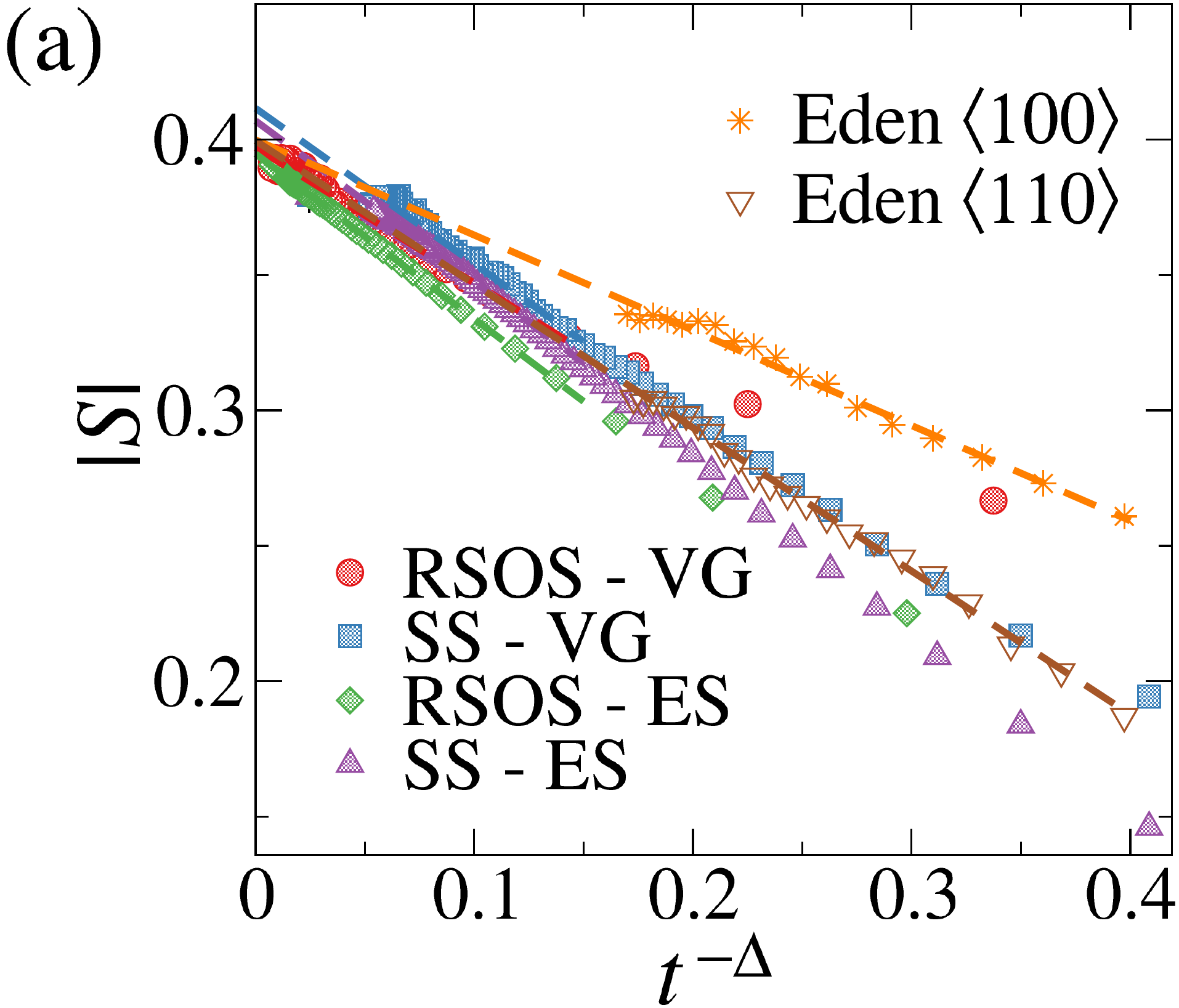}
	\includegraphics[height=3.6cm]{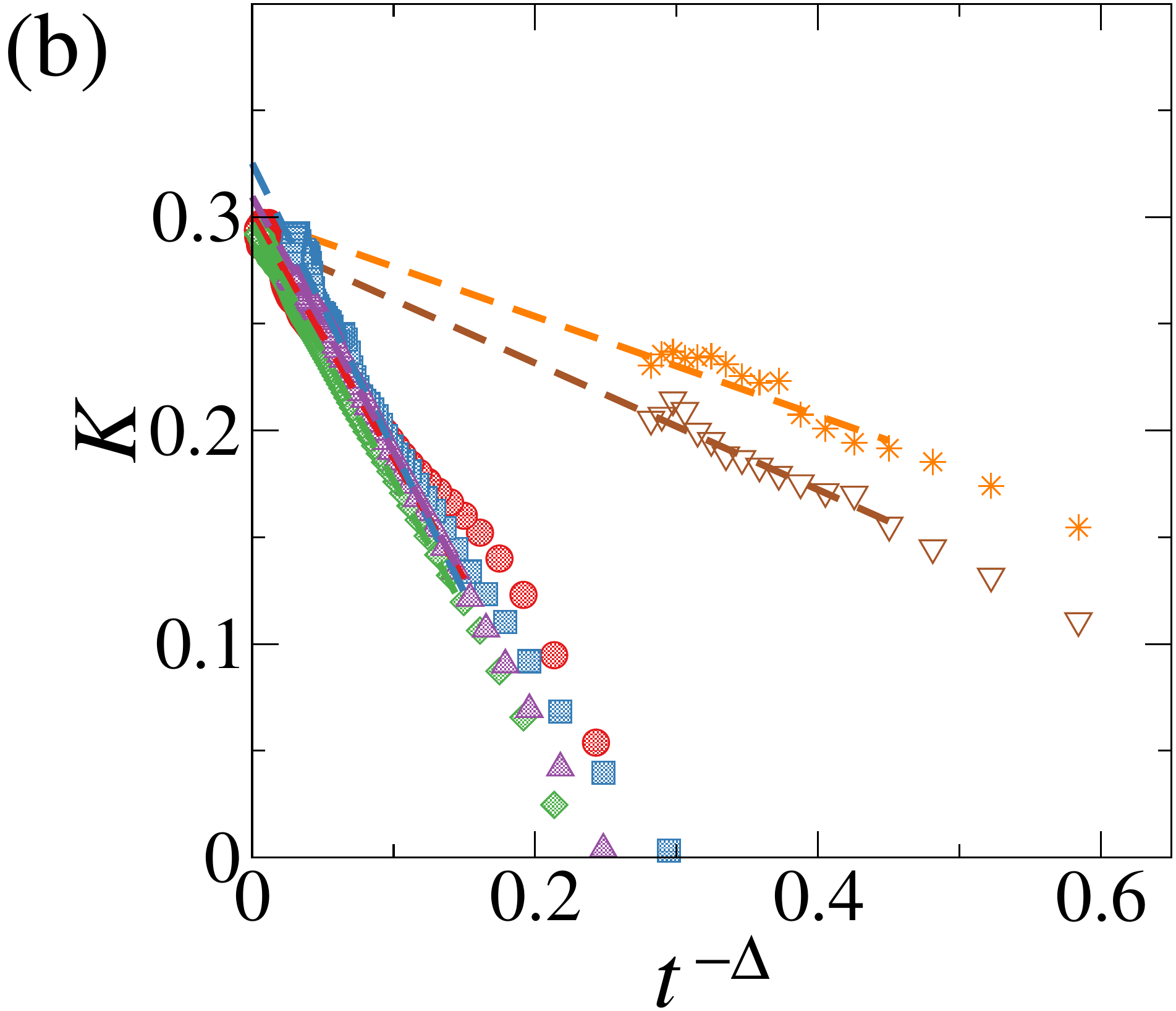}
	\includegraphics[height=3.6cm]{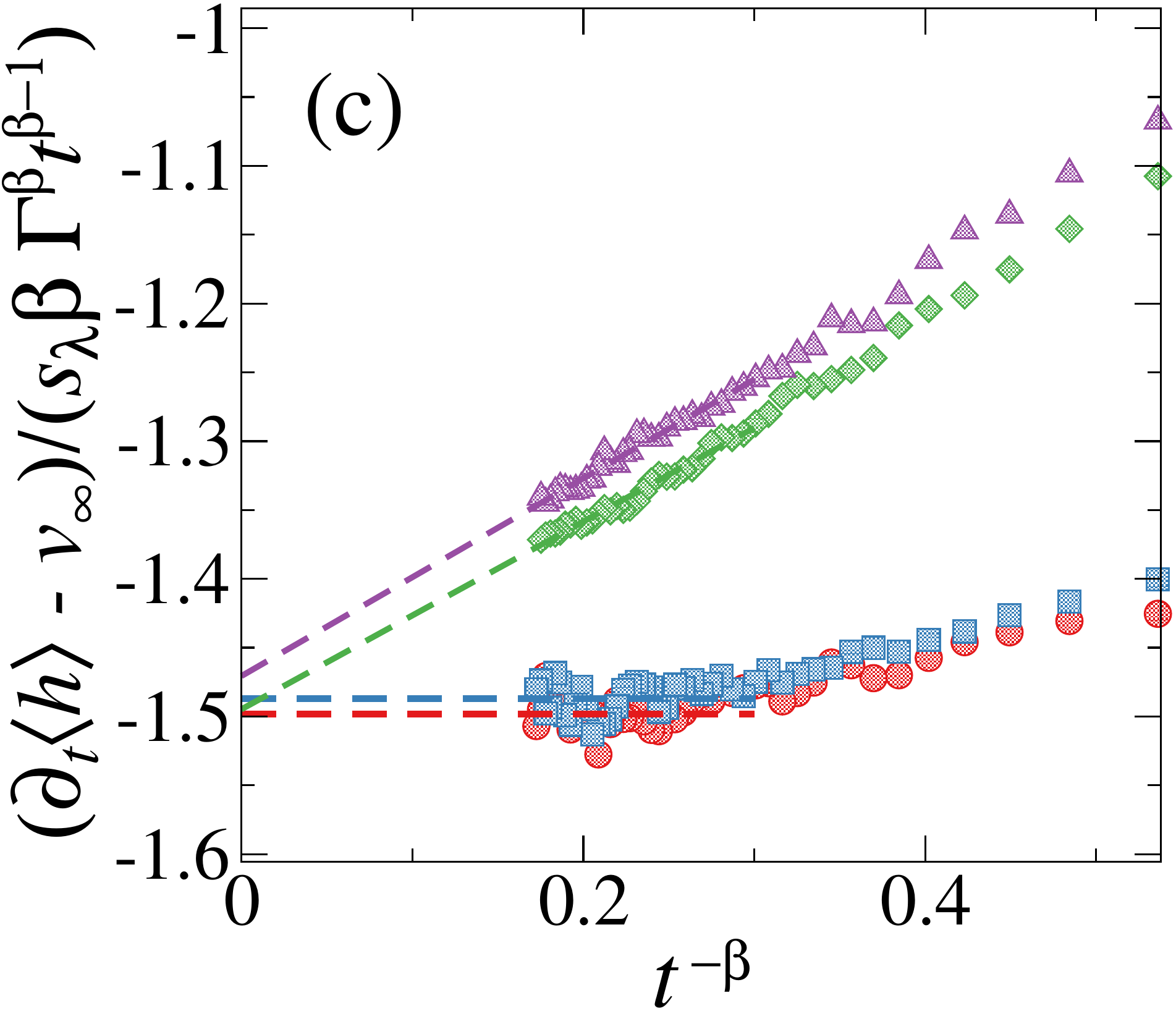}
	\includegraphics[height=3.6cm]{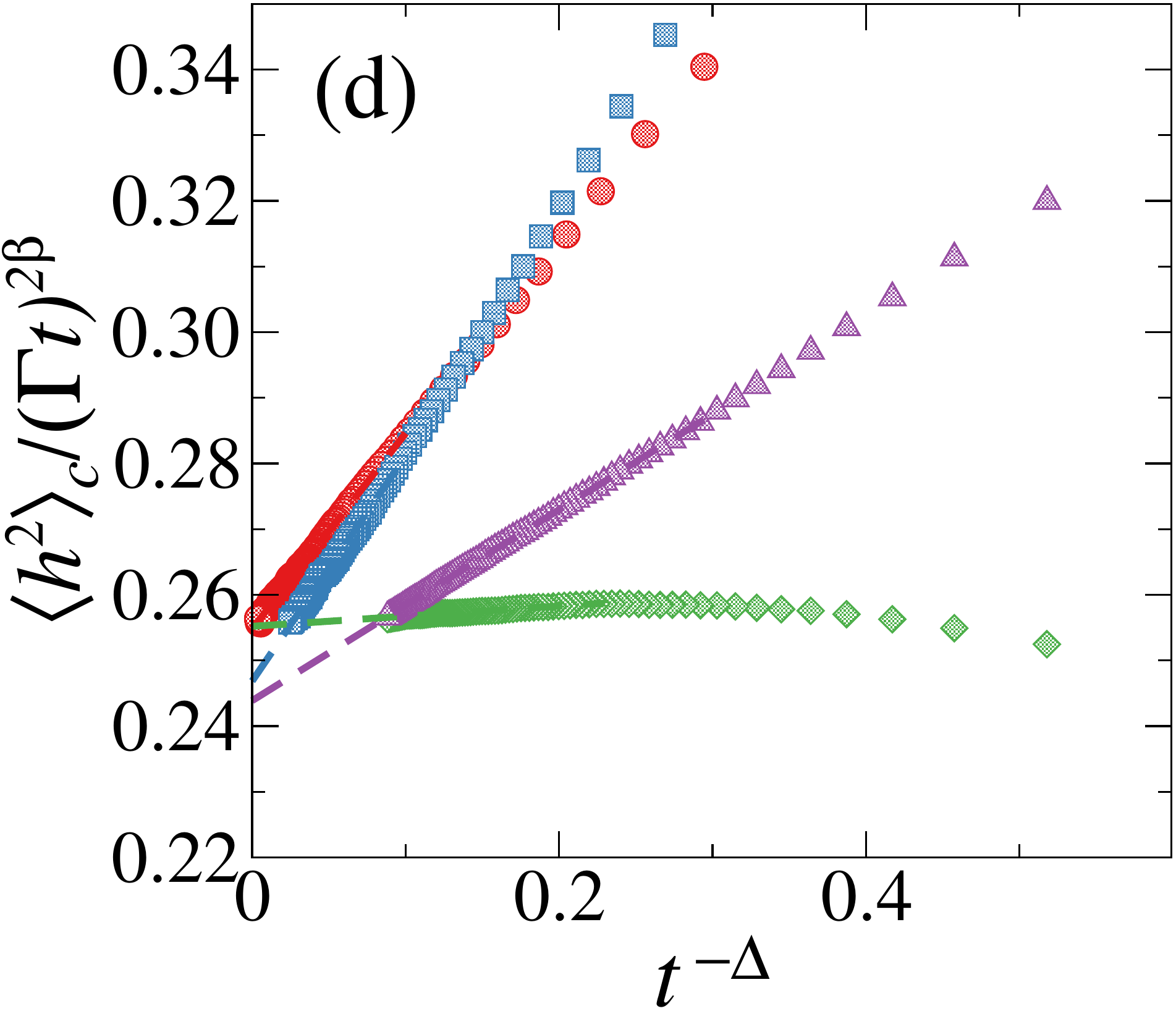}
	\caption{HDs' skewness $|S|$ (a) and kurtosis $K$ (b), and rescaled growth velocity $(\partial_t\langle h\rangle-v_\infty)/(s_{\lambda}\beta\Gamma^\beta t^{\beta-1})$ (c) and variance $\langle h^2\rangle_c/(\Gamma t)^{2\beta}$ (d) versus $t^{-\Delta}$, for the models/setups indicated by the legends in (a). The dashed lines are linear fits used to extrapolate the data to the $t\rightarrow\infty$ limit. In (a) $\Delta=0.33$, $0.66$ and $0.4$ for the Eden, RSOS and SS models, respectively; while in (b) $\Delta=0.25$, $0.75$ and $0.5$ respectively for the same models. The exponents used in (d) were $\Delta=0.75$, $0.5$ and $0.33$ for the RSOS-VG, SS-VG, and both models in ES case, respectively. In panel (c), $\Delta=\beta$ for all systems.}
	\label{Fig3}
\end{figure}

Even more compelling evidence that all systems analyzed here belong to the same universality class is provided in Figs. \ref{Fig3}(a) and \ref{Fig3}(b), where extrapolations to the $t \rightarrow\infty$ limit of the skewness $S=\langle h^3\rangle_c/\langle h^2\rangle_c=\langle \chi^3\rangle_c/\langle \chi^2\rangle_c^{3/2} $ and (excess) kurtosis $K=\langle h^4\rangle_c/\langle h^2\rangle_c^2=\langle \chi^4\rangle_c/\langle \chi^2\rangle_c^2$ of the HDs are, respectively, shown. In fact, the extrapolated values of each of these ratios are very close for all investigated systems, demonstrating that they have the same asymptotic HD. By considering the average and standard deviations of such extrapolated values, we obtain $|S|=0.402(6)$ and $K=0.30(1)$. These results agree quite well with $S=0.40(1)$ and $K=0.31(2)$, corresponding to the averages of the data reported by Halpin-Healy in Table III of Ref. \cite{healy13}, mostly from models for directed polymers in random media (DPRM) with point-line boundary conditions. This confirms the universality of such cumulant ratios for the KPZ HD for ($2+1$) cylindrical geometry.

To fully characterize the first HDs' cumulants, we also have to determine the average $\langle \chi\rangle$ and variance $\langle\chi^2\rangle_c$ of the pdf $P(\chi)$. According to the ansatz in Eq. \ref{eqAnsatz}, to access these quantities, firstly we need to know the nonuniversal parameters $v_{\infty}$ and $\Gamma$ for each model and growth setup. As discussed in Refs. \cite{healy12,healy13}, these parameters are not expected to change with the geometry of the system, so values obtained for flat substrates are expected to also hold in the spherical and cylindrical cases. Indeed, it was observed in Ref. \cite{Ismael14} that such parameters, for the RSOS and SS models, are the same when they are deposited on 2D flat substrates with $L_y = L_z = const.$ (flat) or $\langle L_y \rangle = \langle L_z \rangle \sim t$ (spherical geometry). Therefore, the values $v_\infty=0.31270(1)$ and $\Gamma=0.68(6)$ for the RSOS, and $v_\infty=0.341368(3)$ and $\Gamma=1.2(1)$ for the SS model, as estimated in Ref. \cite{Ismael14} and references therein, will be used for these models here. The consistency of the data obtained below confirms that such values also hold in the cylindrical case. Due to the anisotropy in the Eden clusters, the nonuniversal parameters may depend on the radial direction considered and, thus, the logic of using the parameters for the flat case does not apply here. Hence, we will estimate $\langle \chi\rangle$ and $\langle\chi^2\rangle_c$ considering only the RSOS and SS models. Then, using these cumulants (i.e., assuming their universality), we estimate the nonuniversal parameters for the Eden clusters along the $\langle 100 \rangle$ and $\langle 110 \rangle$ planes in Appendix \ref{secApendA}.

With the values of $v_\infty$ and $\Gamma$ at hand, we can obtain $\langle\chi\rangle $, according to Eq. \ref{eqAnsatz}, by extrapolating $(\partial_t\langle h\rangle-v_\infty)/(s_{\lambda}\beta\Gamma^\beta t^{\beta-1})$ to the $t \rightarrow \infty$ limit. Figure \ref{Fig3}(c) shows such extrapolations, where one sees that the data for the VG case converge quite fast, indicating the existence of weak corrections in the ansatz \ref{eqAnsatz} for these systems, while important corrections exist for the expanding substrates. Despite this, in all cases, the outcomes from the extrapolations are very similar, yielding $\langle \chi\rangle=-1.49(1)$. From Eq. \ref{eqAnsatz} the squared roughness is given by $\langle h^2\rangle_c \simeq (\Gamma t)^{2\beta}\langle\chi^2\rangle_c$, meaning that $\langle\chi^2\rangle_c$ can be obtained by extrapolating $\langle h^2\rangle_c/(\Gamma t)^\beta$ to $t \rightarrow \infty$. These extrapolations are shown in Fig. \ref{Fig3}(d), from which we obtain the average result $\langle \chi^2\rangle_c=0.251(7)$. This value, as well as the one for $\langle\chi\rangle$, are both in good agreement with the previous estimates from Ref. \cite{healy13} [$\langle \chi\rangle=-1.47(2)$ and $\langle \chi^2\rangle_c=0.249(5)$], confirming their universality. 

\begin{table}[!b]
	\caption{Mean $\langle\chi\rangle$, variance $\langle\chi^2\rangle_c$, their ratio $R=\langle\chi\rangle/\sqrt{\langle\chi^2\rangle_c}$, skewness $S$ and kurtosis $K$ of the $(2+1)$ KPZ HDs $P(\chi)$ for the three main geometries. The data for the cylindrical case were obtained by averaging the ones found here with those from Ref. \cite{healy13}. The results for the flat geometry are averages of the data reported in Refs. \cite{healy12,tiago13}, while those for the spherical case were extracted from Refs. \cite{healy13,tiago13,Ismael14} \footnote{$\langle\chi\rangle$ and $\langle\chi^2\rangle_c$ were not reported in Ref. \cite{tiago13}.}.}
	\begin{center}
		\begin{tabular}{l c c c c  c c c c c c }
			\hline \hline
			Geometry & & $\langle\chi\rangle$ & & $\langle\chi^2\rangle_c$ && $R$ && $|S|$ && $K$ \\
			\hline
			Flat        && -0.75(5) && 0.234(15)  && -1.55(15)  && 0.424(6) && 0.345(8)   \\
			Cylindrical && -1.48(3) && 0.250(6)   && -2.96(8)   && 0.401(8) && 0.31(2)  \\
			Spherical   && -2.32(6) && 0.335(15)  && -4.0(2)  && 0.329(10)  && 0.211(6)   \\
			\hline\hline
		\end{tabular}
		\label{tabHD}
	\end{center}
\end{table}

Table \ref{tabHD} presents a summary of these cumulants and the ratios $S$ and $K$, considering the values estimated here and in previous works \cite{healy12,healy13}, comparing them with the averages of estimates for such quantities in the literature for flat and spherical geometries \cite{healy12,healy13,tiago13,Ismael14}. Note that, as expected, the results for the cylindrical case have intermediate values, between those for flat and spherical systems. However, while $\langle\chi^2\rangle_c$, $|S|$ and $K$ for spherical geometry present an appreciable difference from the others, the differences between the values for the flat and cylindrical cases are only $\sim 10$\%. This means that very accurate estimates for these quantities are needed to distinguish between these two geometries, which may be hard to obtain, e.g., in experiments. Thereby, a better adimensional ratio for this matter is $R = \langle\chi\rangle/\sqrt{\langle\chi^2\rangle_c}$ [i.e., the inverse of the coefficient of variation of $P(\chi)$], whose values are quite different for each geometry (see Tab. \ref{tabHD}). Note, however, that $v_{\infty}$ is required to access $R$, since $s_{\lambda}[\langle h\rangle (t) - v_{\infty}t]/\sqrt{\langle h^2\rangle_c(t)} \rightarrow R$ as $t \rightarrow \infty$. This contrasts with $S$ and $K$, which can be estimated without any knowledge of the nonuniversal parameters.

\begin{figure}[!t]
	\centering
	\includegraphics[width=8.cm]{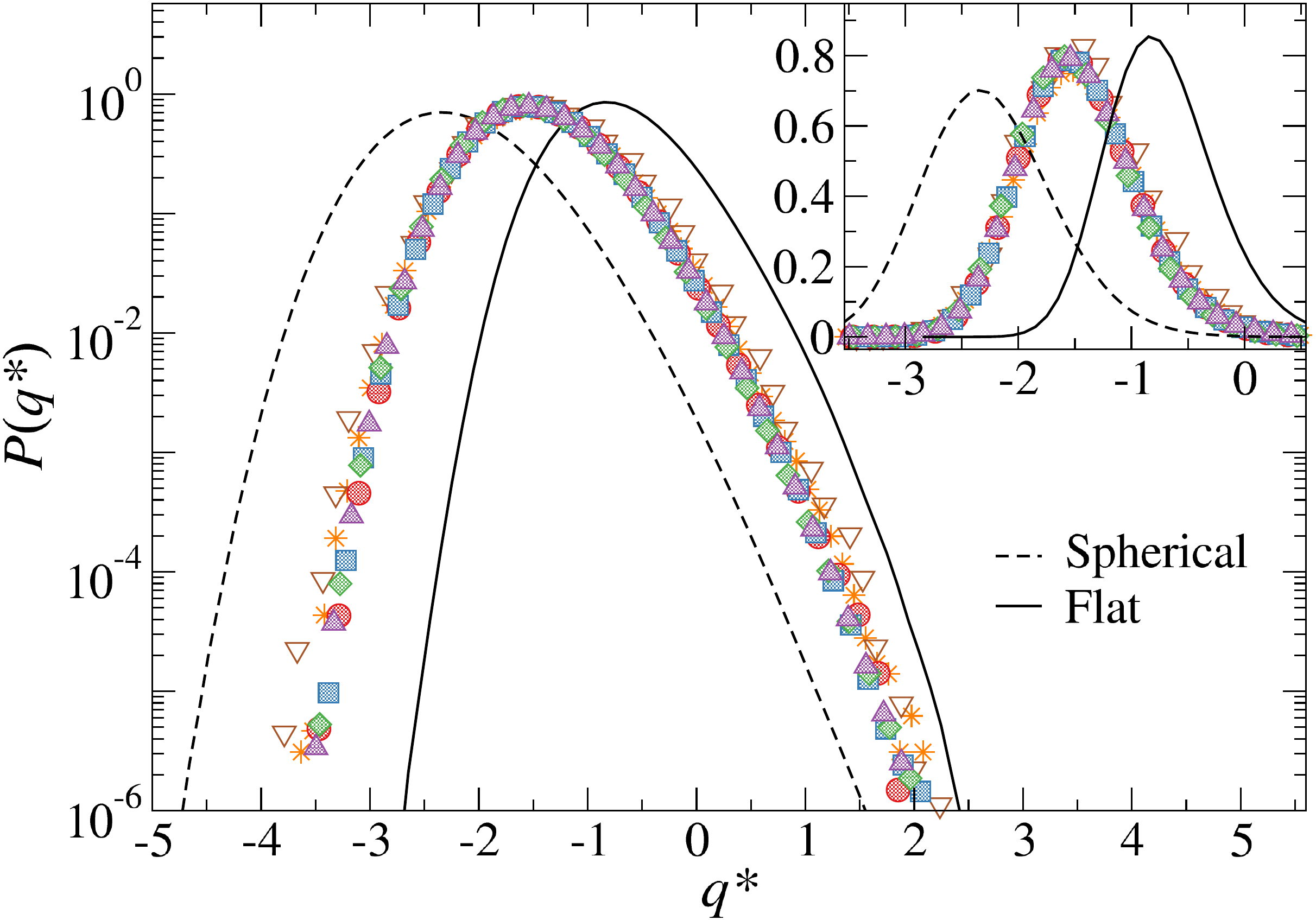} 
	\caption{Rescaled HDs for all models/setups yielding cylindrical growth (symbols). The inset shows the same data as the main plot in linear-linear scale, highlighting the peaks' behavior. The data shown here follow the same symbol and color schemes of Figs. \ref{Fig2} and \ref{Fig3}. The $(2+1)$ KPZ HDs for flat and spherical geometries, obtained from simulations of the RSOS model in Ref. \cite{Ismael14}, are also shown for comparison.}
	\label{Fig4}
\end{figure}

We end this section comparing the pdf's of the HDs for the different models, measured at the longest simulation times in each case. According to Eq. \ref{eqAnsatz}, we could access the distribution of $P(\chi)$ by rescaling the HDs as $q=(h-v_\infty t)/[s_{\lambda}(\Gamma t)^\beta]$ and $P(q)=P(h)(\Gamma t)^\beta$. It turns out, however, that important finite-time corrections exist in the ansatz \ref{eqAnsatz}, as already noticed above, which can not be disregarded in this analysis. Therefore, we will investigate the rescaled HDs $P(q^*)$, where $q^*=[h-v_\infty t-f(t)]/[s_{\lambda}(\Gamma t)^\beta]$ and $f(t)$ accounts for the relevant corrections in each model and growth setup. A detailed analysis of such corrections is presented in Appendix \ref{secApendB}, where one finds that $f(t)$ has a logarithmic behavior in most cases, besides a constant term (see Tab. \ref{tabCor}). Figure \ref{Fig4} presents the rescaled HDs for the different models, where a striking data collapse is observed, providing additional confirmation of their universality. For the sake of comparison, the rescaled HDs for flat and spherical geometries --- obtained from simulations of the RSOS model on substrates with $L_z=L_y = const.$ and $\langle L_z \rangle = \langle L_y \rangle \sim t$, respectively, in Ref. \cite{Ismael14} --- are also shown in Fig. \ref{Fig4}, demonstrating the clear geometry-dependence of the pdf's $P(\chi)$.

\section{Results for the covariances}
\label{secResCov}

Now we investigate the two-point correlators of the 2D KPZ interfaces with cylindrical geometry. We start with the spatial covariance, defined in Eq. \ref{eqCovS}, and then analyze the temporal covariance from Eq. \ref{eqCovT}.

\subsection{Spatial covariances}

 As discussed in the Introduction, the scaling function $\Psi[s]$ (see Eq. \ref{eqCovS}) is expected to assume universal, but different forms in each geometry. This has been indeed demonstrated in numerical works for the spherical \cite{Ismael14} and flat \cite{healy14exp,Ismael14} geometries, as well as experimentally in the flat case \cite{healy14exp,Almeida17}. We remark that in these two geometries the correlations parallel to the interface spread equally in both substrate directions, so $C_S(r,t)$ is the same if one measures it along the $y$ or $z$ directions, or along a circle of radius $r$. In the cylindrical case, on the other hand, the expansion of a single interface side breaks this symmetry, so that different functions $\Psi_y[s]$ and $\Psi_z[s]$ can be obtained by measuring the covariance along \textit{lines} in the azimuthal ($y$) and longitudinal ($z$) directions. Therefore, we will analyze each of these directions separately here. Given the lack of translation invariance in the $y$ direction of the interfaces obtained in the grooved substrates, as well as in the Eden clusters, we are able to explore the azimuthal covariances only in the ES systems. We also notice that, due to the small times attained in the Eden simulations, large deviations (from the rest) are found in its longitudinal covariances. For this reason, these results for the Eden clusters will be omitted here.

\begin{figure}[!t]
	\centering
	\includegraphics[width=4.2cm]{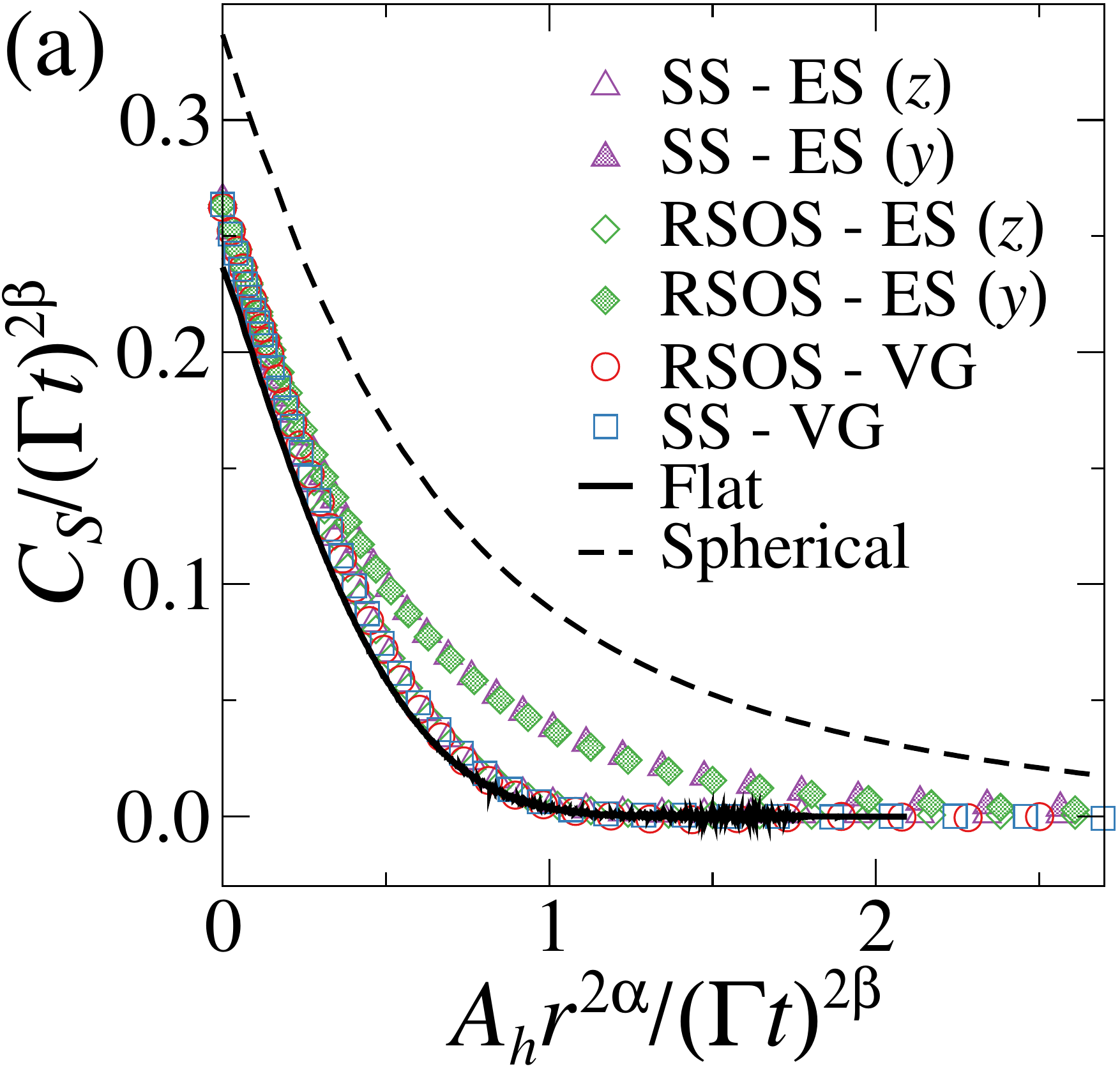}
	\includegraphics[width=4.2cm]{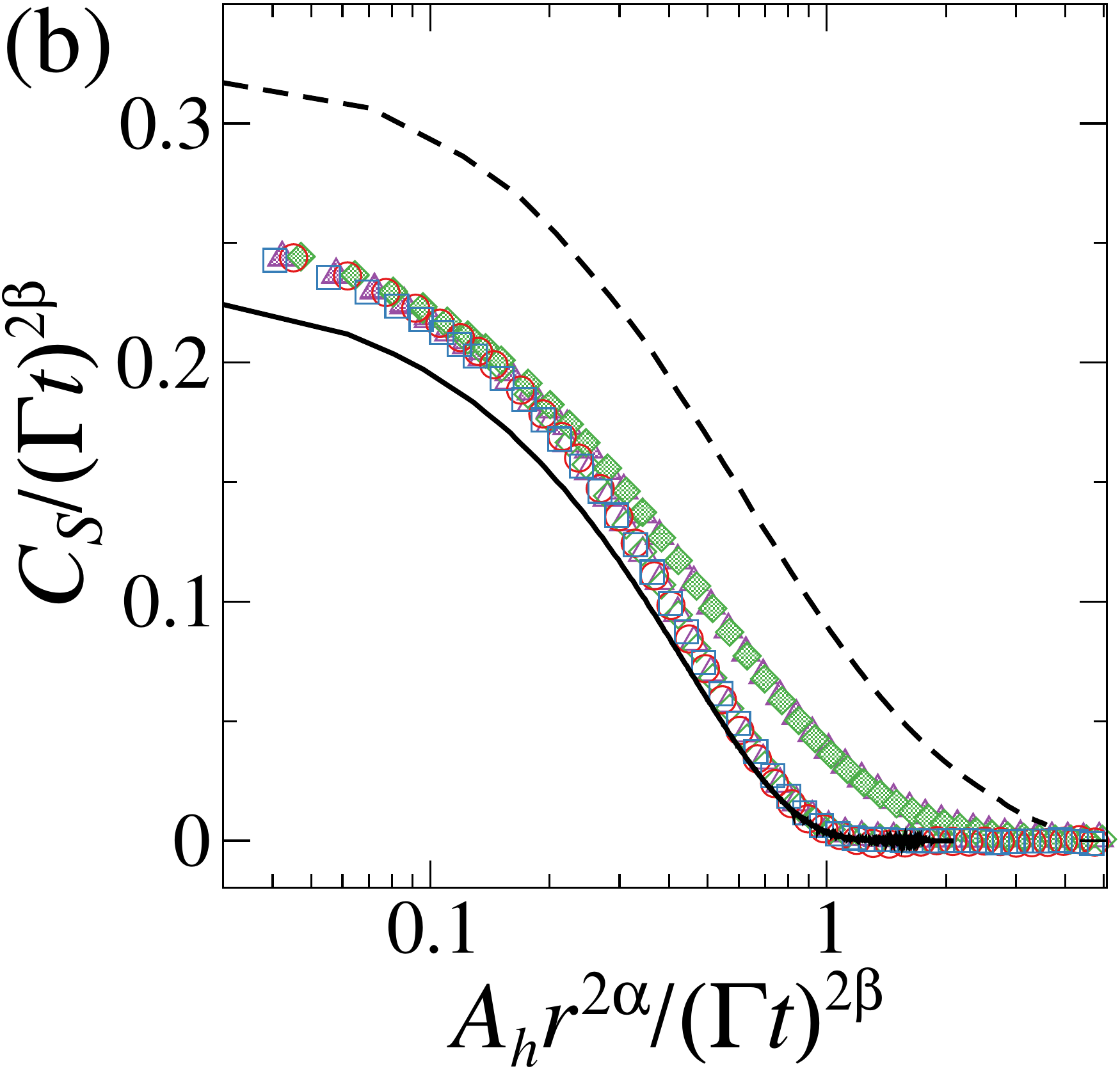}
	\includegraphics[width=4.2cm]{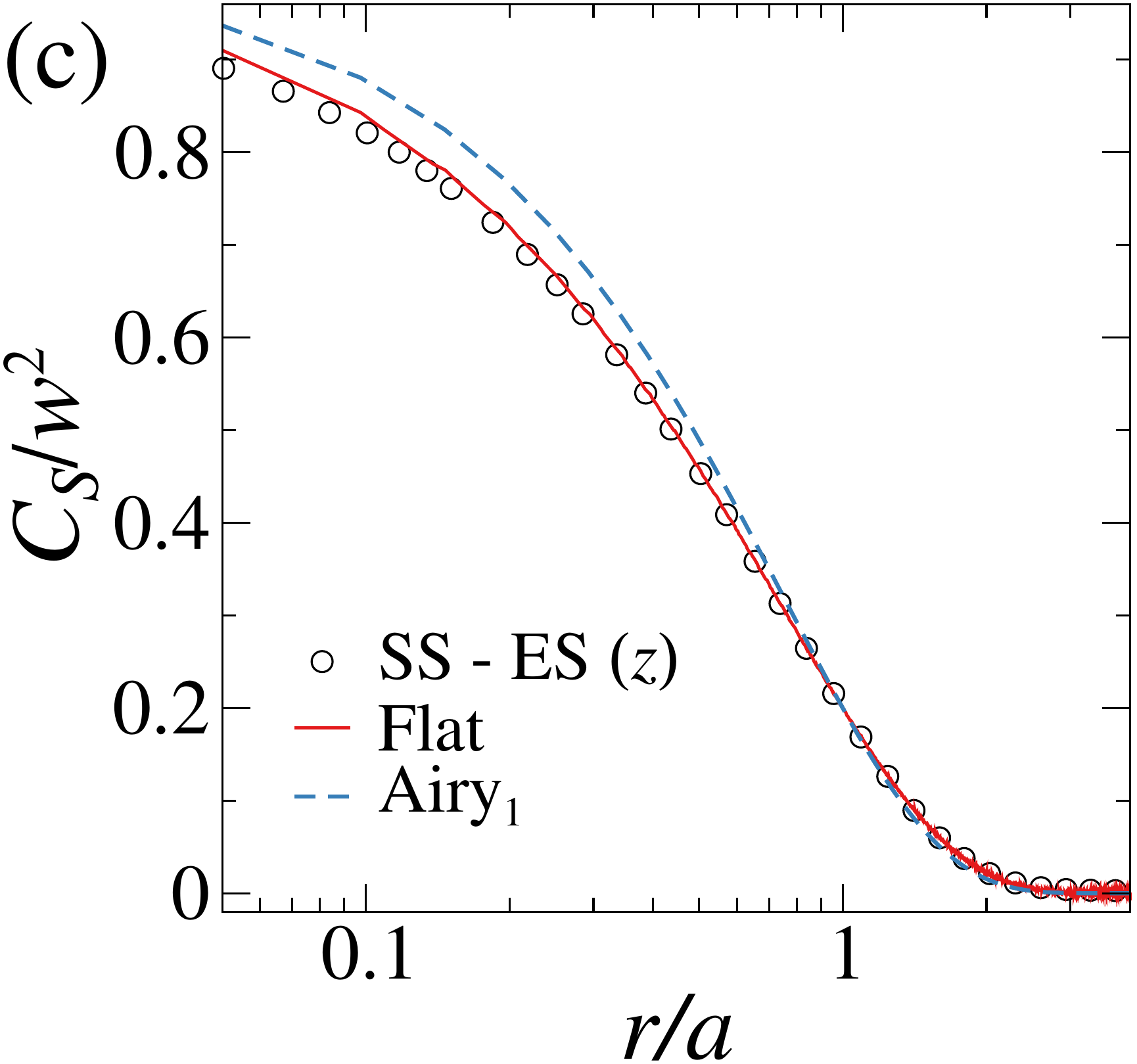}
	\includegraphics[width=4.2cm]{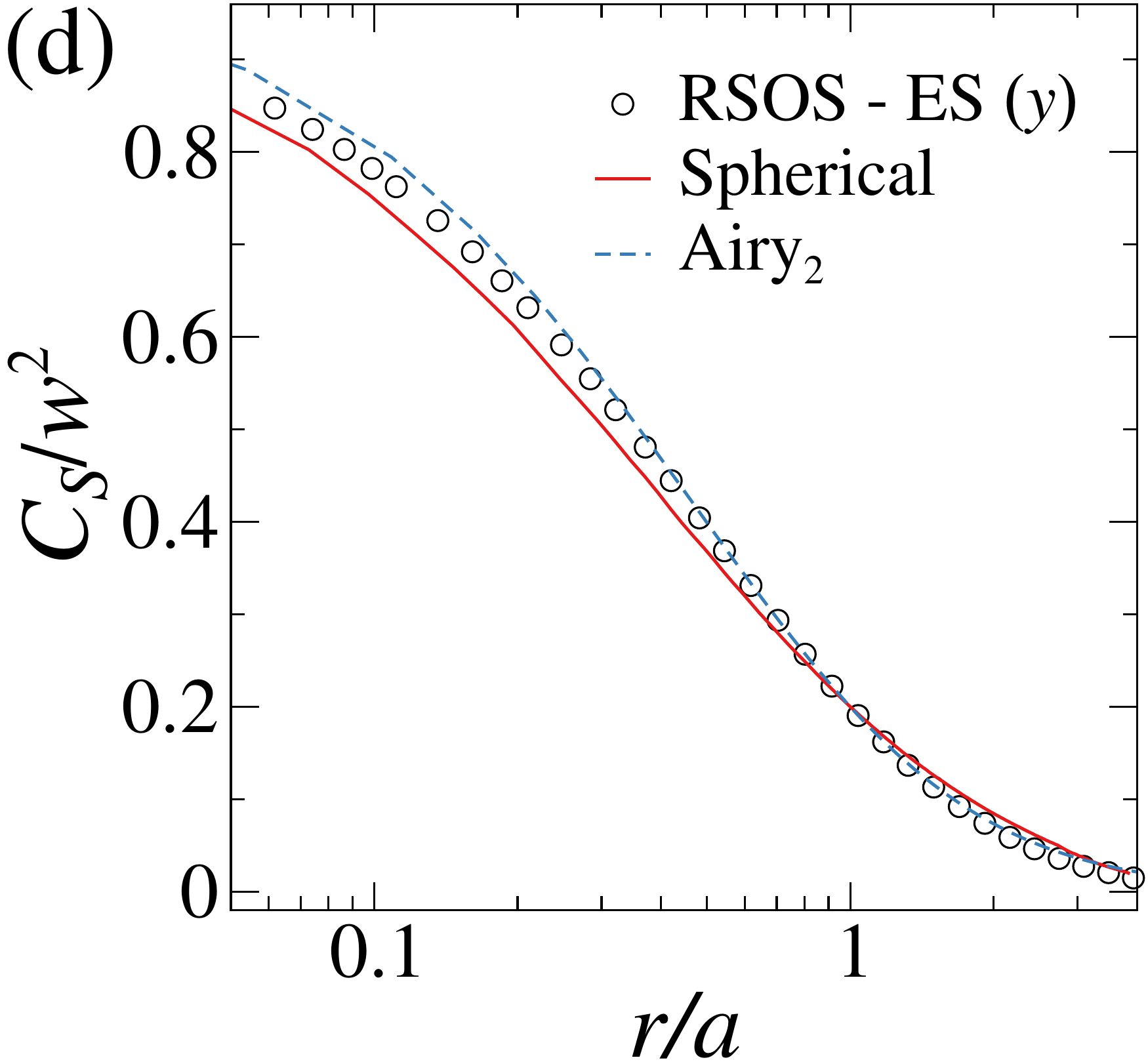}
	\caption{(a) and (b) Rescaled spatial covariance $C_S/(\Gamma t)^{2\beta}$ against $A_h r^{2\alpha}/(\Gamma t)^{2\beta}$; the only difference between these panels is the logarithmic scale in the abscissa in (b). Results for the longitudinal ($z$) [azimuthal ($y$)] direction are represented by open [closed] symbols. (c) and (d) Rescaled spatial covariances $C_S/w^2$ versus $r/a$. In (c), the curve for the $z$ direction (represented by the SS-ES data) is compared with those for flat geometry, in both 1D (Airy$_1$) and 2D (Flat). An analogous comparison, between the curve for $y$ direction (represented by the RSOS-ES data) and those for spherical geometry in 1D (Airy$_2$) and 2D (Spherical), is presented in (d). The results for ($2+1$) flat and spherical systems were numerically obtained for the RSOS model in Ref. \cite{Ismael14}.}
	\label{Fig5}
\end{figure}

To unveil the forms of the scaling functions $\Psi_y[s]$ and $\Psi_z[s]$, we analyze the rescaled curves of $C_S/(\Gamma t)^{2\beta}$ versus $A_h r^{2\alpha}/(\Gamma t)^{2\beta}$ for both directions, which are compared in Figs. \ref{Fig5}(a) and \ref{Fig5}(b). There, the open and closed symbols represent the longitudinal and azimuthal directions, respectively, and one can see that the data from different models and growth setups collapse very well, for a given direction, demonstrating the universality of these covariances. However, as expected, different functions $\Psi_y[s]$ and $\Psi_z[s]$ exist for each direction. Indeed, although both functions start at the same value at the origin, since $C_S/(\Gamma t)^{2\beta} = \langle \chi^2\rangle_c$ at $r = 0$ --- and there exists a single HD $P(\chi)$ for these cylindrical systems ---, they have a very different behavior for large $r$, namely, while the longitudinal correlator saturates at $\Psi_z[s] \approx 0$ for $A_h r^{2\alpha}/(\Gamma t)^{2\beta} \gg 1$ (corresponding to $r \gg \xi$), in the expanding direction $\Psi_y[s]$ is always a decreasing function of $s$. These behaviors are analogous to those found for the covariance of flat and spherical 2D KPZ interfaces, respectively, as can be seen in Figs. \ref{Fig5}(a) and \ref{Fig5}(b), where the curves for these two geometries are also shown for comparison. We may note in these figures that $\Psi_z[s]$ is somewhat close to the covariance of the flat case, but it lays between the curves for flat and spherical systems, as clearly observed in Fig. \ref{Fig5}(b), where the abscissa is presented in logarithmic scale. Such in-betweeness of the cylindrical covariances is even more evident in $\Psi_y[s]$. 

It may be argued that this simple comparison of the $\Psi$'s for different geometries is not so fair because they have different values of $\langle \chi^2\rangle_c$, starting thus at different points. Therefore, to better compare the forms of the scaling functions, we analyze them under the rescaling $G[u]=C_S/w^2$ versus $u=r/a$, where $a$ is the value that forces the curves to pass at $C_s/w^2=0.2$ at $r/a=1$. In this way, all curves have the same value at $r/a=1$ and also at $r/a=0$, where $C_S/w^2=1$. Figures \ref{Fig5}(c) and \ref{Fig5}(d) show such curves, where, for the sake of clarity, we only present data for a single model in each direction, in the cylindrical case. This makes it quite evident that the curve for the longitudinal direction, $G_z[u]$, is indeed very similar to the one for the flat geometry, since they agree very well for $r/a \gtrsim 0.4$, having only a slight deviation for smaller $u$ [see Fig. \ref{Fig5}(c)]. On the other hand, we may observe in Fig. \ref{Fig5}(d) that the curve for the azimuthal direction does not agree with the one for the spherical case neither for small nor large $r/a$. We also present in these figures the curves for flat and circular 1D KPZ interfaces, i.e., the Airy$_1$ and Airy$_2$ processes, respectively. While the Airy$_1$ curve is quite different from $G_z[u]$ for small $u$, the difference is not so large between the Airy$_2$ and $G_y[s]$ curves, particularly for $r/a \gtrsim 0.5$. This indicates that, considering a single line in the direction with fixed size (i.e., the longitudinal direction in the cylindrical KPZ system or any direction in the flat case), the 2-pt correlations on it are not so sensitive to what is happening in the other direction. On the other hand, the correlations measured along an expanding line presents a stronger variation depending on whether the other direction is fixed or expanding.

\subsection{Temporal covariance}

As noticed in Sec. \ref{secIntro}, the rescaled temporal covariance $\mathcal{A}(t/t_0)$ in Eq. \ref{eqCovT} scales as $\mathcal{A}(t/t_0) \sim (t/t_0)^{-\bar{\lambda}}$ when $t \gg t_0$. With bases on exact calculations for linear growth equations \cite{KrugKallabis97} and a simple geometric argument, Kallabis and Krug \cite{Kallabis99} conjectured that $\bar{\lambda} = \beta + d/z$ is valid in general for flat interfaces of dimension $d$. This was indeed verified for 1D \cite{Kallabis99,Takeuchi2012,Ismael14} and 2D KPZ systems \cite{Ismael14}, as well as for the Villain-Lai-Das Sarma (VLDS) \cite{Villain,*LDS} class in $d=1$ and $2$ \cite{Ismael16a}. The explanation for this behavior is as follows \cite{Kallabis99}: ``$\mathcal{A}(t/t_0)$ measures the overlap of the height configurations at times $t$ and $t_0$, and this overlap is the product two factors: (\textit{i}) the lateral overlap between domains in the $d$-dimensional substrate space, which is of order $[\xi(t_0)/\xi(t)]^d \sim (t/t_0)^{-d/z}$; and (\textit{ii}) the horizontal overlap $W(t_0)/W(t) \sim (t/t_0)^{-\beta}$.''

For hyperspherical linear interfaces, however, Singha \cite{Singha2005} demonstrated that $\bar{\lambda} = \beta$. Relying on simulations and analytical approaches for KPZ systems, it was conjectured in Ref. \cite{Singha2005} that $\bar{\lambda} = \beta$ is valid in general for radially growing systems. This has been indeed confirmed for the KPZ \cite{Singha2005,Takeuchi2012,Ismael14} and VLDS \cite{Ismael16a} classes. This behavior can be understood in light of the reasoning above (by Kallabis and Krug \cite{Kallabis99}), by noting that in expanding systems the lateral overlap [factor (\textit{i})] becomes irrelevant, so that the decay is ruled out solely by the horizontal contribution (\textit{ii}).

By the same token, if $d^*$ of the $d$ interface directions are kept fixed during the growth, while the other $(d-d^*)$ expand radially, the lateral overlap shall matter only for the $d^*$ ones, leading to the generalized exponent:
\begin{equation}
 \bar{\lambda} = \beta + \frac{d^*}{z}.
 \label{ConjecCT}
\end{equation}
Therefore, in flat geometry, where all lateral substrate sizes are fixed, one has $d^*=d$, as proposed in Ref. \cite{Kallabis99}. In the hyperspherical case, on the other hand, all directions expand, so $d^*=0$, yielding the Singha's result \cite{Singha2005}. Importantly, in the cylindrical systems analyzed here, $d^*=1$ and, then, assuming that $\beta \approx 0.241$ (as discussed above) and then $z=2/(\beta+1) \approx 1.611$, one obtains $\bar{\lambda} \approx 0.862$. We notice that, considering the rational exponents proposed in Ref. \cite{tiago22} ($\beta=7/29$ and $z=29/18$) in Eq. \ref{ConjecCT}, one gets $\bar{\lambda} = (7+18 d^*)/29$ for 2D KPZ systems, which gives $\bar{\lambda} = 25/29 = 0.86206$ in the cylindrical case. 

\begin{figure}[!t]
	\centering
	\includegraphics[width=8cm]{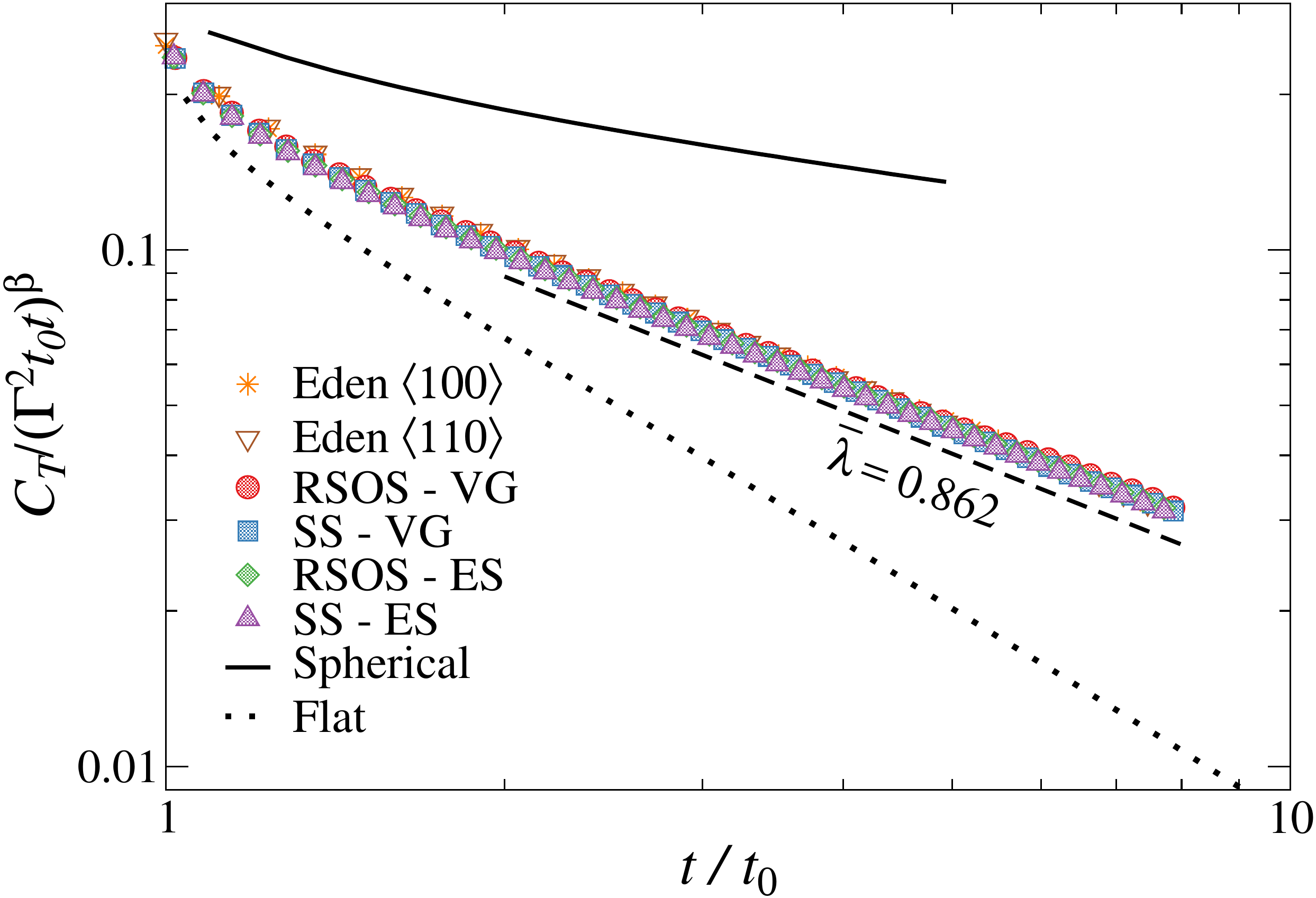} 
	\caption{Rescaled temporal covariance $C_T/(\Gamma^2t_0t)^\beta$ against the ratio $t/t_0$, for the ($2+1$) cylindrical systems indicated by the legend (symbols). The dashed line has the indicated slope. The initial times are $t_0=25$ for the Eden, and 187 and 204 for the other models in ES and VG cases, respectively. Numerical estimates of the curves $\mathcal{A}(t/t_0)$, obtained for the ($2+1$) RSOS model in Ref. \cite{Ismael14} for spherical and flat geometries, are also shown for comparison.}
	\label{Fig6}
\end{figure}

Figure \ref{Fig6} shows the rescaled temporal covariances for all models analyzed here, where a striking data collapse is observed, demonstrating that the finite-time corrections in this quantity are much weaker than those observed above in the HDs and spatial covariances. Substantially, for long times one finds a behavior consistent with $\mathcal{A}(t/t_0) \sim (t/t_0)^{-0.862}$, strongly indicating that our conjecture for the exponent $\bar{\lambda}$ in Eq. \ref{ConjecCT} is correct. 

For comparison, Fig. \ref{Fig6} presents also the rescaled covariances for $(2+1)$ KPZ systems with flat and spherical geometries (as obtained in Ref. \cite{Ismael14} from simulations of the RSOS model on substrates with $L_y=L_z=const.$ and $\langle L_y \rangle=\langle L_z \rangle \sim t$, respectively). As expected, the curves are quite different for each geometry, since they decay asymptotically with the exponents $\bar{\lambda}_{spher.} \approx 0.241$, $\bar{\lambda}_{cylind.} \approx 0.862$ and $\bar{\lambda}_{flat} \approx 1.482$.

\section{Summary}
\label{secConc}

We have presented a thorough study of the universal properties of the statistics of the ($2+1$) cylindrical KPZ subclass. In all investigated systems, the roughness scale with growth exponents in agreement or very close to the value expected for 2D KPZ systems, confirming that these models belong to the KPZ class for all growth setups considered here. The cumulants (and their adimensional ratios) obtained here for the HDs agree quite well among the different models, as well as with those reported in Ref. \cite{healy13}, providing solid evidence of the universality of this limit distribution. In this context, we emphasize that the HDs' skewness and kurtosis for the flat and cylindrical cases are somewhat close, so a better ratio to distinguish between these HDs is the inverse of the coefficient of variation, $R = \langle\chi\rangle/\sqrt{\langle\chi^2\rangle_c}$, since it has quite different values in each geometry, as demonstrated in Tab. \ref{tabHD}.

The spatial covariances have different forms when measured in the longitudinal and azimuthal directions, with the curve for the former case being very similar to the one for $(2+1)$ flat KPZ systems. The curve for the azimuthal direction resembles the Airy$_2$ covariance for circular 1D KPZ interfaces, when appropriately rescaled. This demonstrates how important the role of the interface expansion is during the growth process, since it not only changes the statistics of systems with different geometries (or ICs), but it can yield different spatial correlators even in the same interface, if it expands anisotropically. Hence, the spatial covariance is more sensitive to details of the system than the HDs, as expected and recently observed also in the context of KPZ systems that expand isotropically but nonlinearly in time \cite{Ismael19,Ismael22}.

For the temporal covariance, previous conjectures for its long time decay in flat and spherical geometries \cite{Kallabis99,Singha2005} were generalized here, and we argue that the related exponent is $\bar {\lambda} = \beta-d^*/z$, with $d^*$ being the number of interface sides whose size is kept fixed during the growth. This is quite well supported by the numerical results for the $(2+1)$ cylindrical systems. Thinking of the aging dynamics of KPZ systems (and other interface growth as well), one expects that $C_T(t,t_0)=t_0^{-2\beta}F_C(t/t_0)$, where $F_C(y) \sim y^{-\lambda_C/z}$ \cite{Daquila,Henkel,OdorAging,healy14exp,Kelling2018}. Therefore, our conjecture implies that the autocorrelation exponent is $\lambda_C = d^*$. While a mathematically rigorous confirmation of this behavior for the KPZ class is a difficult task, demonstrating this for linear growth equations (e.g., the Edwards-Wilkinson \cite{EW} and Mullins-Herring ones \cite{Mullins,*Herring1951}) is an interesting project.

Unveiling the complete picture of the KPZ statistics for higher dimensions is another interesting point to be tackled in the future. Although the HDs for flat geometry have been analyzed in some recent works up to $d=6$ \cite{Alves14,HHTake2015,Alves16,Kim2019}, it seems that no result exists for other geometries and nothing is known for the spatial covariances. The results here and elsewhere for $d=2$ strongly suggest that $d+1$ different limit HDs and temporal covariances shall exist, depending on whether the radial growth starts from a single seed (i.e.,  the hyperspherical case, where $d^*=0$), from a seed line (hypercylindrical case, where $d^*=1$), and so on until the flat case (where $d^* = d$). Moreover, an even larger number of spatial covariances (measured along the various substrate directions, as done here) are expected.

\acknowledgments

The authors acknowledge financial support from CNPq and FAPEMIG (Brazilian agencies).

\appendix

\section{Non-universal parameters of the cylindrical Eden clusters}
\label{secApendA}

\begin{figure}[!b]
	\centering
	\includegraphics[width=4.25cm]{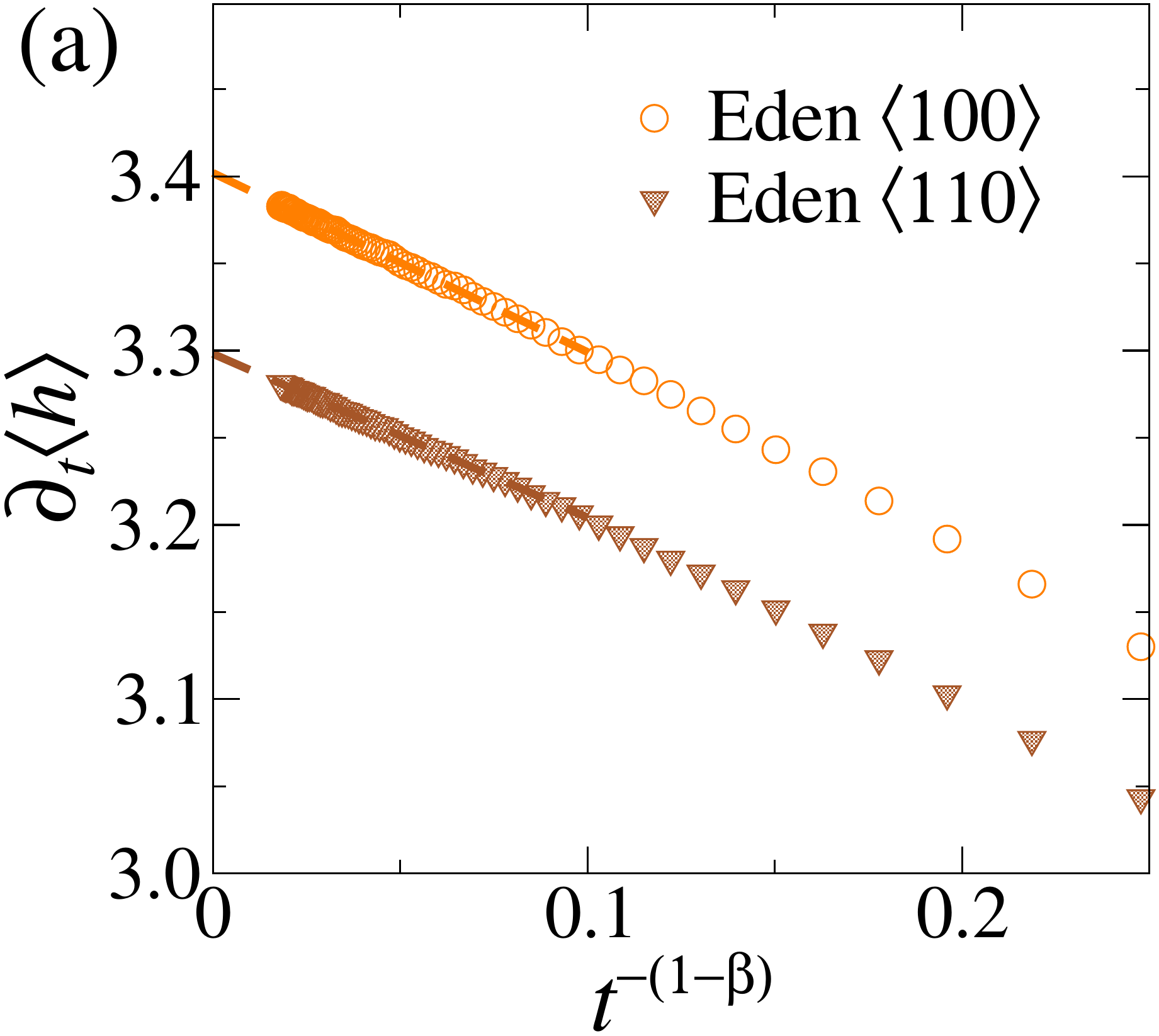}
	\includegraphics[width=4.25cm]{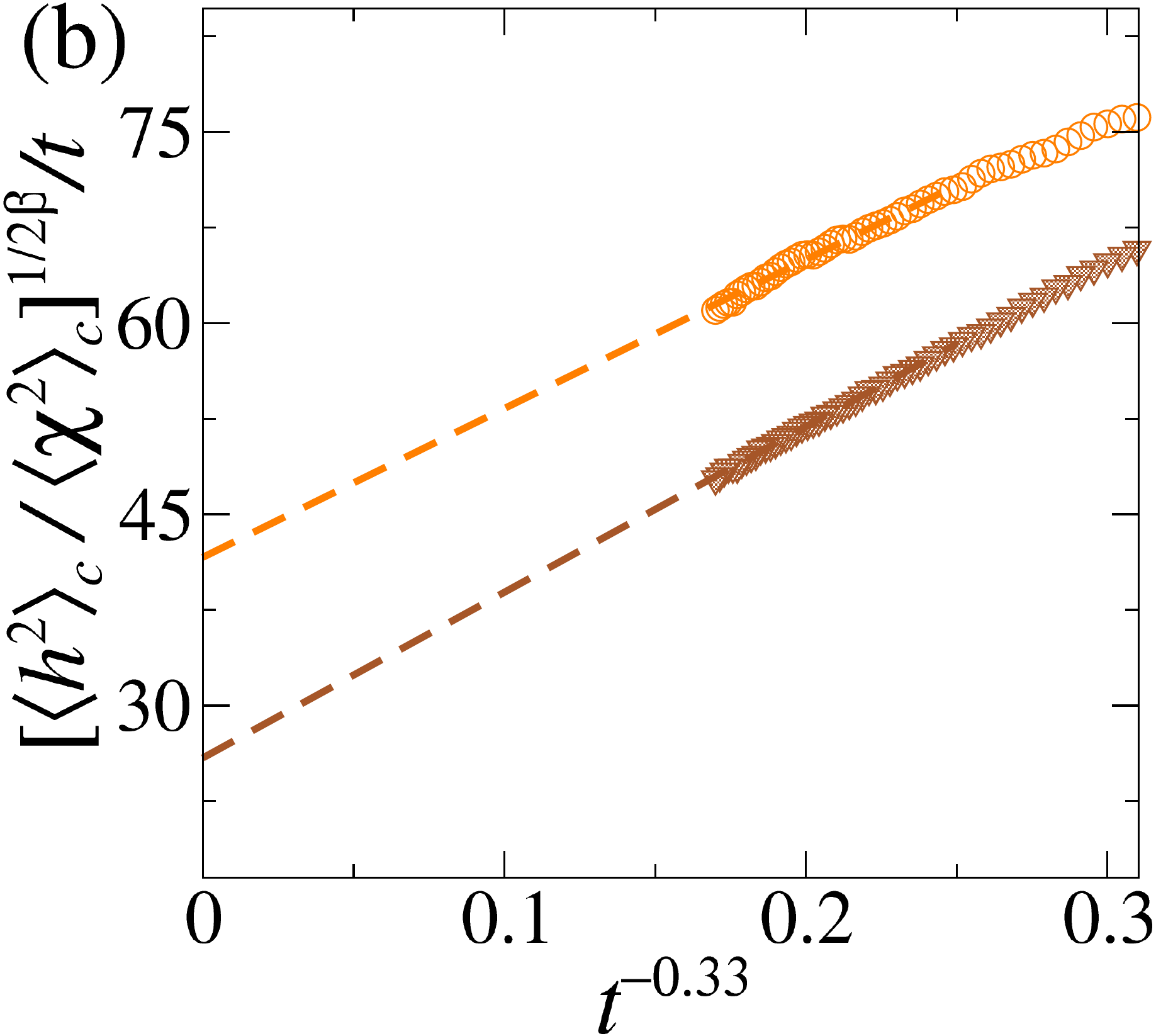}
	\caption{(a) Growth velocity $\partial_t \langle h \rangle$ versus $t^{-(1-\beta)}$ and (b) rescaled HDs' variance $[\langle h^2\rangle_c/\langle \chi^2\rangle_c]^{1/2\beta}/t$ as a function of $t^{-0.33}$, for the cylindrical Eden clusters, measured along lines parallel to the $z$ axis in the lattice planes (Eden ${\langle 100\rangle}$) and in the diagonal planes (Eden ${\langle 110\rangle}$). The dashed lines are linear fits used to extrapolate the data to $t\rightarrow\infty$.}
	\label{FigApendA}
\end{figure}

Eden clusters growing radially on the lattice acquire anisotropic shapes because each radial direction evolves with a different growth velocity $v(t) = \partial_t \langle h \rangle$. This is confirmed in Fig. \ref{FigApendA}(a), which shows $v(t)$ measured in lines parallel to the $z$ axis (i.e., the longitudinal direction of the cylinder) on the $xz$ and $yz$ planes, denoted by $v_{\langle 100 \rangle}$, and on the diagonal planes, $v_{\langle 110\rangle}$. From Eq. \ref{eqAnsatz}, one expects that $v(t) = v_{\infty} + b t^{-(1-\beta)}$ and indeed $v(t)$ versus $t^{-(1-\beta)}$ presents a good linear behavior at long times, as demonstrated in Fig. \ref{FigApendA}(a). The extrapolations of these data to the $t \rightarrow \infty$ limit return the asymptotic growth velocities $v_{\langle 100 \rangle,\infty}=3.4020(6)$ and $v_{\langle 110 \rangle,\infty}=3.2984(8)$. As an aside, we note that, as expected, these velocities are considerably larger than those reported in Ref. \cite{Alves13} for two-dimensional Eden clusters grown from a single seed on the square lattice, which are $v_{\langle 10 \rangle}=2.1824(3)$ and $v_{\langle11\rangle}=2.1401(3)$ for the lattice and diagonal directions, respectively.

To obtain the parameter $\Gamma$, we consider the second cumulant of the HDs, which (according to Eq. \ref{eqAnsatz}) is expected to behave asymptotically as $\langle h^2\rangle_c \simeq (\Gamma t)^{2\beta}\langle \chi ^2\rangle_c$. Hence, we estimate $\Gamma$ by extrapolating $[\langle h^2\rangle_c/\langle \chi^2\rangle_c]^{1/2\beta}/t$ to the $t\rightarrow\infty$ limit, considering that $\langle \chi^2\rangle_c=0.250(6)$, as estimated for the other models in Sec. \ref{secResWHDs}. Figure \ref{FigApendA}(b) shows such extrapolations, which yield $\Gamma_{\langle 100\rangle}=41(3)$ and $\Gamma_{\langle 110\rangle}=26(2)$. Despite the strong finite-time corrections in this figure, the nice collapses observed in Figs. \ref{Fig4}, \ref{Fig5} and \ref{Fig6} between the data for the Eden and the other models confirm the reliability of these estimates for the non-universal parameters of the cylindrical Eden clusters in both directions considered. (Note that other values for $v_{\infty}$ and $\Gamma$ should be found if one had analyzed them for other radial directions.)

\section{Corrections in the KPZ ansatz}
\label{secApendB}

The KPZ ansatz, as presented in Eq. \ref{eqAnsatz}, is valid asymptotically (i.e., for $t \rightarrow \infty$), while important finite-time corrections may exist at short times. As observed in the experiments by Takeuchi \textit{et al.} \cite{Takeuchi2010,Takeuchi2011}, as well as in subsequent theoretical and numerical works (see, e.g., Refs. \cite{Sasamoto2010,tiago12a,tiago13,Ismael14}), one such correction is an additive variable $\eta$. Moreover, evidence that another additive term of type $\zeta t^{-a}$ is also present has been reported in some works \cite{tiago12a,tiago13,Ismael14}. Hence, one may expect that
\begin{equation}
 h \simeq v_{\infty} t + s_\lambda (\Gamma t)^{\beta} \chi + \eta + \zeta t^{-a},
\label{eqAnsatzCor}
\end{equation}
where $a>0$ is a constant, and $\eta$ and $\zeta$ are, in principle, stochastic variables. Note that these additional terms are considered corrections because they both become irrelevant in $\langle h \rangle$, as well as in $\langle h^n \rangle_c$ as $t\rightarrow\infty$. 

\begin{figure}[!t]
	\centering
	\includegraphics[height=3.6cm]{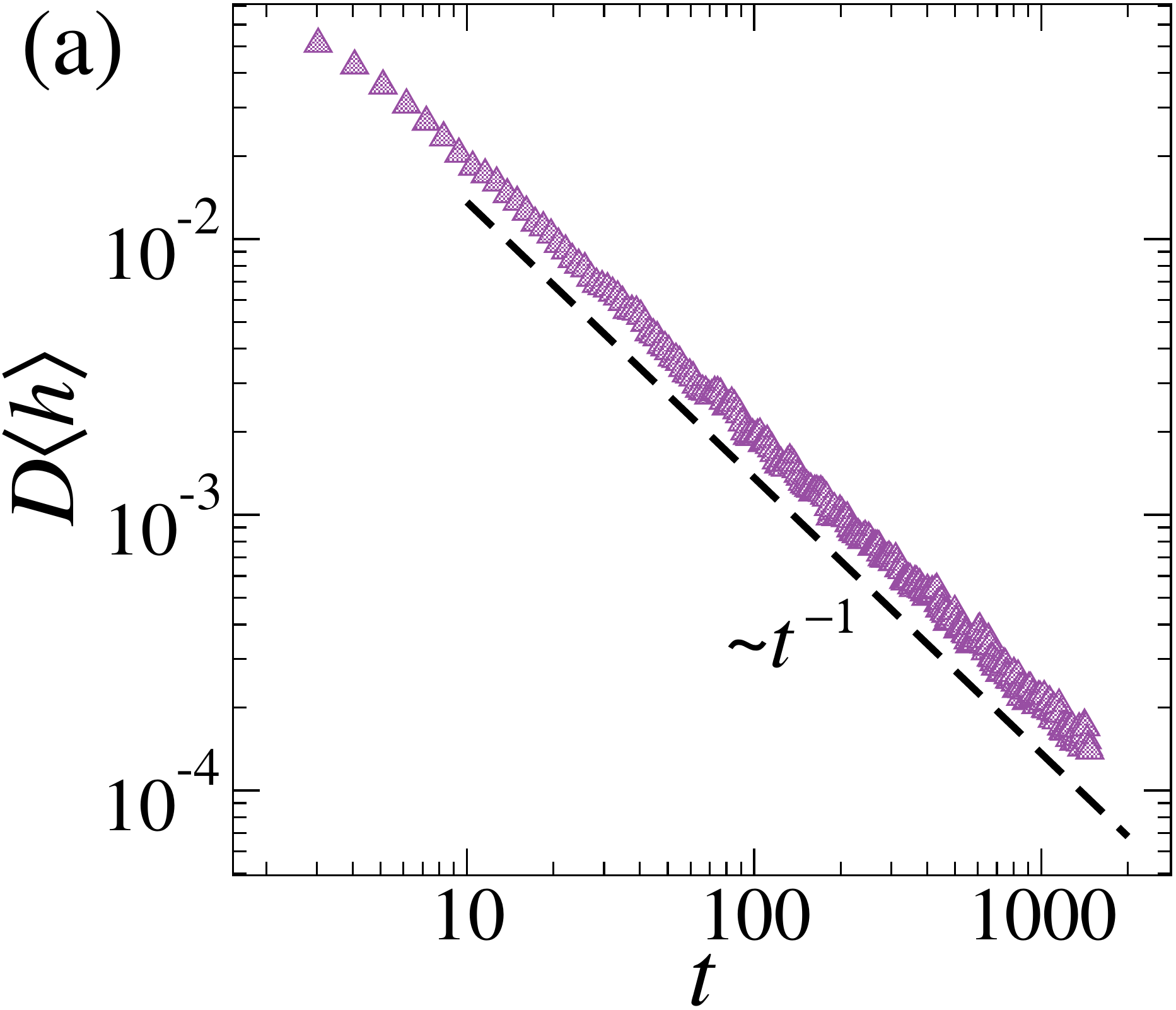}
	\includegraphics[height=3.6cm]{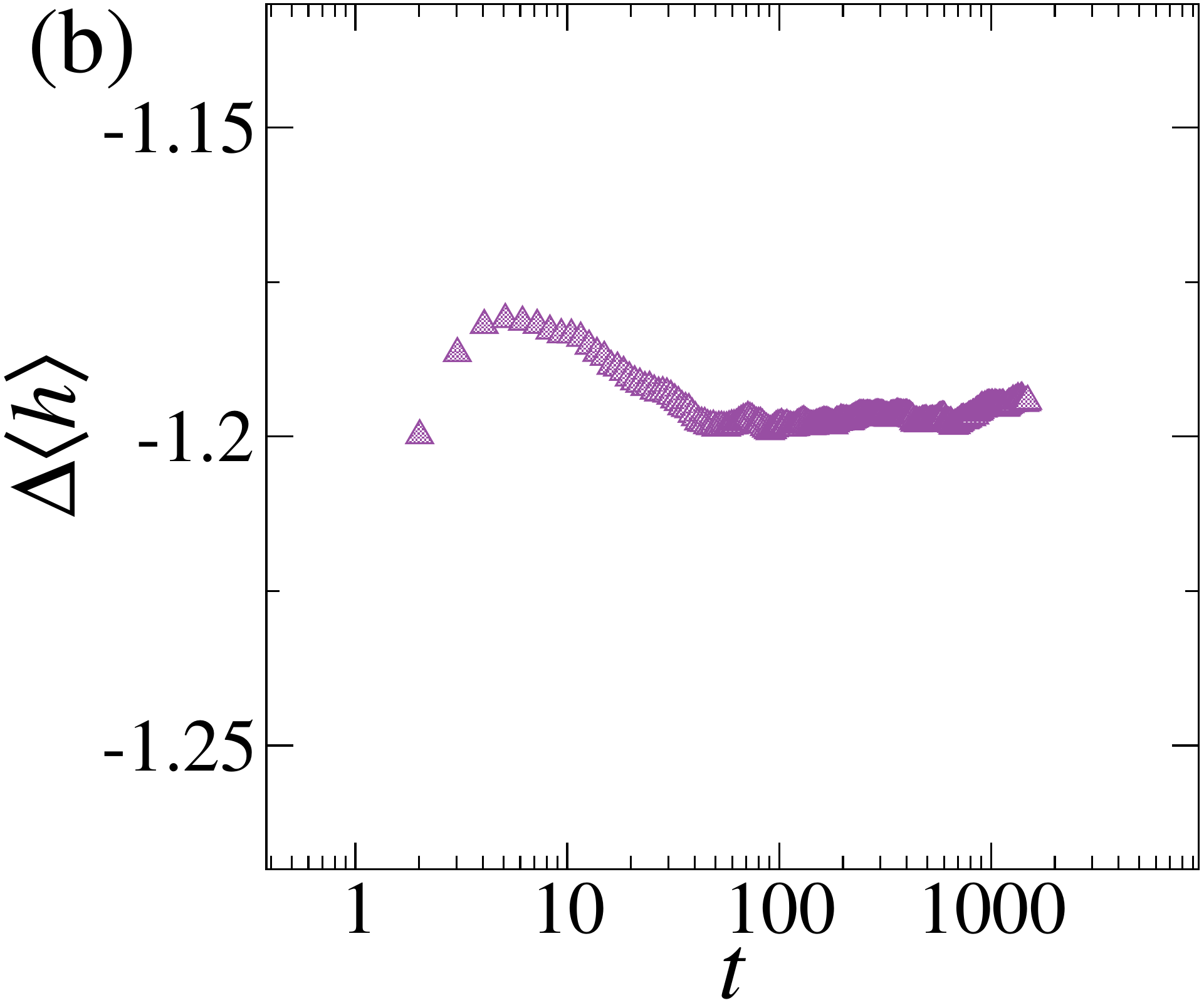}
	\caption{Temporal evolutions of the differences (a) $D \langle h\rangle$ and (b) $\Delta \langle h\rangle$ (see the text for definitions) for the SS model in the ES setup. The dashed line in (a) has the indicated slope.}
	\label{FigApendB}
\end{figure}

\begin{table}[!b]
	\caption{Corrections in the KPZ ansatz for the different models and growth setups.}
	\begin{center}
		\begin{tabular}{l c c c c }
			\hline \hline
			System & & &&$f(t)$ \\
			\hline
			Eden $\langle 100 \rangle$ &&& & $+2.0(1)-0.42(3)\ln(t)$ \\
			Eden $\langle 110 \rangle$ &&& & $+1.15(9)-0.35(4)\ln(t)$ \\
			RSOS - ES                  &&& & $-1.45(2) -0.15(1)\ln(t)$ \\
			SS - ES                    &&& & $-1.20(3)-0.18(1)\ln(t)$ \\
			RSOS - VG                  &&& & $-1.10(2)-0.18(6)t^{-0.39(7)}$ \\
			SS - VG                    &&& & $-0.84(1)-0.40(5)t^{-0.55(8)}$ \\
			\hline\hline
		\end{tabular}
		\label{tabCor}
	\end{center}
\end{table}

To estimate these corrections, it is convenient to start from the last term in Eq. \ref{eqAnsatzCor}, since $\langle \eta \rangle$ can be eliminated by a derivative of $\langle h\rangle$ with respect to time, so $D\langle h\rangle \equiv \partial_t\langle h\rangle-v_\infty - s_{\lambda} \beta \Gamma^\beta t^{\beta-1}\langle \chi\rangle \simeq -a \langle \zeta \rangle t^{-a-1}$. Therefore, we may estimate $a$ and $\langle \zeta \rangle$ from a power law fit in a log-log plot of $D\langle h \rangle$ versus $t$. Figure \ref{FigApendB}(a) shows an example of such a plot for the SS model in the ES setup, where one finds that $D\langle h \rangle$ decays consistently with $\sim t^{-1}$. This indicates the existence of a logarithmic correction in the ansatz, as indeed expected for KPZ growth on substrates that enlarge linearly in time \cite{Ismael14,Ismael19,Ismael22}. Hence, in this case $D\langle h \rangle \sim \langle \zeta \rangle t^{-1}$ and $\langle \zeta \rangle$ can be obtained from the scaling amplitude in Fig. \ref{FigApendB}(a). Similar logarithmic behaviors are found for the RSOS model in the ES setup, as well as for the Eden models, whereas power-law corrections are obtained in the VG case (see Tab. \ref{tabCor}).

Once the last correction term in Eq. \ref{eqAnsatzCor} is determined, one simply has to plot $\Delta \langle h\rangle \equiv \langle h\rangle-v_\infty t-s_\lambda (\Gamma t)^\beta \langle \chi\rangle - \langle \zeta \rangle t^{-a}$ versus $t$ (with $t^{-a}$ replaced by $\ln t$ when $a=0$) to obtain the value of $\langle \eta \rangle$. Figure \ref{FigApendB}(b) presents this plot for the SS model in the ES case, where one can see a fast convergence of the curve to an approximately constant value. Besides confirming that the logarithmic behavior obtained in Fig. \ref{FigApendB}(a) is correct, this also demonstrates that further corrections in the ansatz are negligible. The temporal average of the curve at the plateau gives us an estimate for $\langle \eta \rangle$.

Table \ref{tabCor} presents a summary of the obtained corrections for each model and growth setup. Note that such corrections were denoted simply as functions $f(t)$ in Sec. \ref{secResWHDs} and, since they were used in the rescale of Fig. \ref{Fig4}, the very good data collapse observed there is a strong evidence of their correctness.

\bibliography{bibExpKPZ2D}

\end{document}